\documentclass{iucr}              
\journalcode{A}
                   \def\href#1{\relax}\let\foo\caption
\ifPDF
  \RequirePackage{hyperref}
  \PassOptionsToPackage{pdftex,bookmarksopen,bookmarksnumbered}{hyperref}
\fi
\let\caption\foo

     \paperprodcode{a000000}      
     \paperref{xx9999}            
     \papertype{FA}               

     \paperlang{english}          
     \journalyr{2018}
     \journalreceived{\relax}
     \journalaccepted{\relax}
     \journalonline{\relax}

\usepackage{amsmath}
\usepackage{amssymb}
\usepackage{mathtools}
\usepackage{graphicx}
\usepackage{oubraces}
\usepackage[colorinlistoftodos]{todonotes}
\usepackage{amsthm}

\DeclareMathOperator{\Tr}{Tr}
\DeclareMathOperator{\Diag}{Diag}
\DeclarePairedDelimiter\abs{\lvert}{\rvert}%

\begin{document}                  



\title{A hidden Markov model for describing turbostratic disorder applied to carbon blacks and graphene}
\shorttitle{possibly a subtitle}


	\cauthor[a]{A. G.}{Hart}{a.hart@bath.ac.uk} \\
     	\author[b]{T. C.}{Hansen} 
     	\author[c]{W. F.}{Kuhs} 
     
        \aff[a]{University of Bath, Bath, UK}
    	\aff[ab]{Institut Laue-Langevin, Grenoble, France}
	    \aff[c]{GZG Abt. Kristallographie, Universit\"{a}t G\"{o}ttingen, Germany}


\shortauthor{Hart, Hansen and Kuhs}





\maketitle                        

\begin{abstract}
We present a mathematical framework to represent turbostratic disorder in materials like carbon blacks, smectites, and twisted $n$-layer graphene. In particular, the set of all possible disordered layers, including rotated, shifted, and curved layers form a stochastic sequence governed by a hidden Markov model. The probability distribution over the set of layer types is treated as an element of a Hilbert space, and using tools of Fourier analysis and functional analysis, we develop expressions for the scattering cross sections of a broad class of disordered materials. 
\end{abstract}


\section{Introduction}

\citeasnoun{warren1941x} was ahead of his time when he observed that certain heat treated carbon blacks appear to comprise
a sequence of equally spaced graphite layers with some random rotation and parallel translation between them. 
In a paper published the next year, \citeasnoun{Warren1942} decided to name this type of disorder among layers \emph{turbostratic disorder}.
The authors used the word to describe
\emph{``graphite layers stacked together roughly parallel and equidistant, but with
each layer having a completely random orientation about the layer normal"} while \citeasnoun{Disorder_first_second} state \emph{``When all the layers are randomly misoriented, the stack is (ideally) turbostratic and there
are no Bragg reflections other than those belonging to the $00l$ series."} 
In the ensuing years, the meaning of the word turbostratic has evolved, with a widely cited paper by \citeasnoun{LI20071686} allowing turbostratic disorder to include layers that are shifted, rotated and curved over some non-uniform probability distribution. This broad definition of the word turbostratic is the definition we will use in this paper, with the goal of bringing together a broad range of disorders under the same mathematical framework.

The notion of turbostratic disorder was developed by \citeasnoun{ShiThesis} to model carbon blacks as a sequence of turbostratically disordered carbon layers that each depend on their preceding layer. Moreover, \citeasnoun{Shihw0018} have written a program CARBONXS  that computes the scattering cross section of their theoretical carbon blacks. The performance of CARBONXS has been compared by \citeasnoun{ZHOU201417} to GSAS, a traditional Rietveld refinement program that does not take turbostratic effects into account, suggesting that an appreciation of turbostratic disorder is necessary to obtain a good fit to X-ray diffraction data.

Since Shi's models are a good fit for carbon blacks, the material appears to truly comprise a Markov chain of turbostratically disordered carbon layers. However, since Shi's model allows carbon layers to be shifted both parallel and perpendicular to the basal plane by any magnitude, there is an uncountable infinity of positions a carbon layer may find itself in. Carbon blacks therefore comprise a Markov chain of layers with infinite state space, and are therefore not fully understood by the analysis of chains with finite state space explored by \citeasnoun{riechers2015pairwise} \citeasnoun{varn2016did} and \citeasnoun{MarkovPaper1}.
The properties of carbon blacks are well worth exploring, as the material has many applications including the moderation of neutrons \cite{ZHOU201417}, lithium-ion batteries \cite{ShiThesis}, and the manufacture of rubber \cite{UNGAR2002929}

\begin{figure}
  \centering
\includegraphics[]{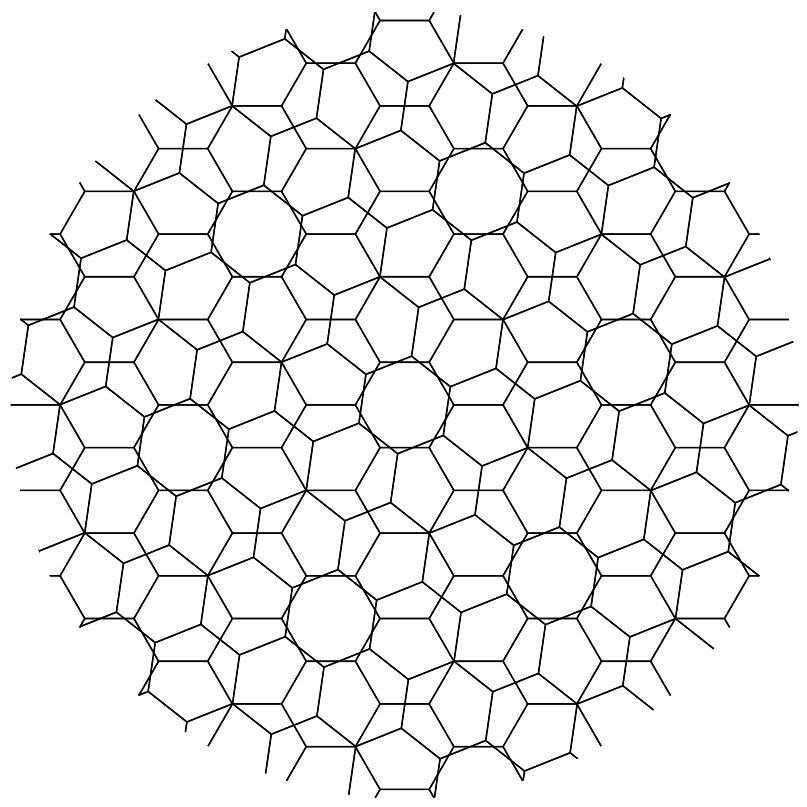}
\caption{An example of a moir\'{e} superlattice with angle $\theta_1 \approx 21.8^\circ$.}
\label{superlattice}
\end{figure}

Aside from carbon blacks, it is now well known \cite{Huang2017} \cite{Razado-Colambo2016} that layers of \emph{graphene} can be stacked atop one another with rotation between them adopting an angle $\theta \in [-\frac{\pi}{6}, \frac{\pi}{6}]$ taking one of an uncountable infinity of values. The rotation can adopt any angle, but the 6 fold rotational symmetry of graphene allows us to work in the restricted range $\theta \in [-\frac{\pi}{6}, \frac{\pi}{6}]$. It is intriguing that a particular countably infinite subset of $[-\frac{\pi}{6}, \frac{\pi}{6}]$ has been the object of great interest and fruitful research among the nanoscience community; specifically the countable set of angles $\theta_i$ for which a pair of layers differing by these angles form a moir\'{e} pattern. Under these angles a twisted bilayer forms a superlattice, and the resultant crystal takes on a so called commensurate structure. 
\citeasnoun{PhysRevLett.99.256802} derived an expression for the moir\'{e} angles $\theta_i$ as exactly the set of angles satisfying
\begin{align}
    \cos(\theta_i) = \frac{3i^2 + 3i + 1/2}{3i^2 + 3i + 1}
\end{align}
for $i = 0, \ 1, \ 2, \ ...$ . An illustration of the superlattice produced when a pair of layers differ by an angle $\theta_1$ is shown in Figure \ref{superlattice}. The electronic properties of moir\'{e} graphene are rich and exotic, and have become subject of a huge international research effort; see for example \citeasnoun{Huang2017}, \citeasnoun{Razado-Colambo2016} \citeasnoun{doi:10.1021/nl204547v}, \citeasnoun{doi:10.1021/nl301137k}, or \citeasnoun{superconducting_angle}, where the most recent authors identified a magic angle where a twisted bilayer becomes a superconductor.  Though the moir\'{e} angles are importantly distinct from the other rotation angles, the latter are still of interest, in fact \citeasnoun{Bistritzer26072011} have remarked that for all other angles $\theta$, a twisted bilayer has no unit cell, but has instead a quasi-periodic structure with its own set of properties.

Most of the relevant nanoscience literature is focused on the simplest interesting model - a single twisted bilayer - but by stacking several layers atop eachother one can form twisted $n$-layer graphene \cite{doi:10.1021/nl301137k}. Assuming any one of the $n$ layers' angle of rotation depends only on a previous layer's angle of rotation, twisted $n$-layer graphene can be described by Markov chain with either a countable or an uncountable number of layer types; depending on whether we insist the rotation angles are moir\'{e}, or allow them to take any value in $[-\frac{\pi}{6},\frac{\pi}{6}]$. In any case, the rotation angles would follow a probability distribution, which has been sought experimentally by \citeasnoun{doi:10.1021/nl204547v} who attempted to infer it from scattering data. Their empirical distribution is compared to torque each atom is subject to, as well as the potential energy per atom.

Smectites (clays) are another class of turbostratically disordered materials. They have been scrutinised under Rietveld refinement by \citeasnoun{Ufer:2008:0009-8604:272} and \citeasnoun{Turbostratic2009} but the turbostratic effects have not been treated rigorously, and may have a more natural description in the framework presented here.

\section{The scattering cross section}

For a crystal composed of otherwise identical layers that differ only by rotation, translation, or change in curvature, the structure factor of the layers are related too. In particular, a layer's structure factor is related by a Fourier transform to the layer's atomic positions, which are related by some rotation, translation or curvature map to the atomic positions of some other layer. To formalise this idea, suppose an arbitrarily chosen reference layer is composed of a periodic array of unit cells. Then for a given unit cell, we express positions in reciprocal space $\vec{Q}$ with reciprocal primitive lattice vectors $\vec{a}^*, \vec{b}^*,$ $\vec{c}^*$ and real numbers $h,k,$ and $l$ such that
\begin{align}
     \vec{Q} = 2 \pi (h \vec{a}^* + k \vec{b}^* + l \vec{c}^*).
\end{align}
Hence, a unit cell comprising $N$ atoms, each with an atomic form factor $f_j$ and position $\vec{r}_j$ with $j = 1 \ ... \ N$ has structure factor $ F_{\text{unit}}$ given by
\begin{align}
    F_{\text{unit}} = \sum_{j=1}^{N}f_{j}e^{-i \vec{Q}\cdot\vec{r}_j}.
\end{align}
The structure factor of an entire layer of unit cells is
\begin{align}
    F = \sum_{(m_1 , m_2) \in \mathcal{D}}\sum_{j=1}^{N}f_{j}e^{-i \vec{Q}\cdot(\vec{r}_j + m_1\vec{a} + m_2\vec{b})} \\ =
    F_{\text{unit}}\sum_{(m_1 , m_2) \in \mathcal{D}}e^{-i(m_1\vec{Q}\cdot\vec{a} + m_2\vec{Q}\cdot\vec{b})}
    \\ =
    F_{\text{unit}}\sum_{(m_1 , m_2) \in \mathcal{D}}e^{- 2 \pi i(m_1 h + m_2 k)}
\end{align}
where $\vec{a}$ and $\vec{b}$ are the primitive lattice vectors that span the basal plane and $\mathcal{D}$ is some subset of $\mathbb{Z}^2$ defining the shape of the layer. If for example the layers are elliptical then $\mathcal{D} \subset \mathbb{Z}^2$ represents some set of lattice nodes enclosed by an ellipse, which we might expect for layers of carbon black crystallites given that \citeasnoun{UNGAR2002929} found the crystallites themselves to be ellipsoidal. If we consider a simpler case of each layer being rectangular with equal dimensions
\begin{align}
\mathcal{D} = [0, ... ,N_a-1] \times [0, ... , N_b-1]
\end{align}
then we obtain the structure factor of a single layer
\begin{align}
    F =
    F_{\text{unit}}\sum_{(m_1 , m_2) \in \mathcal{D}}e^{-2 \pi i(m_1 h + m_2 k)} \\
    = F_{\text{unit}}\sum_{m_1 = 0}^{N_a-1} e^{-2 \pi im_1 h} \sum_{m_2 = 0}^{N_b-1} e^{-2 \pi im_2 k}
    \\
    = F_{\text{unit}}\frac{\sin(N_a \pi h)}{\sin(\pi h)}\frac{\sin(N_b \pi k)}{\sin(\pi k)}e^{-i(N_a-1)\pi h}e^{-i(N_b-1)\pi k}
    \label{structure_factor_of_layer}
\end{align}
where the last line follows from the definition of the Dirichlet kernel.
The contribution of this layer to the scattering pattern $S$ is $\abs{F}^2$ allowing us to recover a perhaps familiar expression
\begin{align}
    S = \abs{F_{\text{unit}}}^2\frac{\sin(N_a \pi h)^2}{\sin(\pi h)^2}\frac{\sin(N_b \pi k)^2}{\sin(\pi k)^2}.
\end{align}
The term
\begin{align}
    \eta(\vec{Q}) = \frac{\sin(N_a \pi h)^2}{\sin(\pi h)^2}\frac{\sin(N_b \pi k)^2}{\sin(\pi k)^2}
\end{align}
is called the shape function, can be modified to represent the different shapes crystallites can take. This is discussed by \citeasnoun{Shihw0018}, \citeasnoun{Warren_text} and \citeasnoun{ERGUN1976139}.

Next, we consider the structure factors of two layers that differ by some rotation. Suppose the rotation is defined by the orthonormal matrix $X$, then we multiply $X$ to the position $\vec{r}_j$ of each atom in the unit cell, as well as the lattice vectors themselves, and find the structure factor $F^X$ of the rotated layer is
\begin{align}
    F^X = \sum_{(m_1 , m_2) \in \mathcal{D}}\sum_{j=1}^{N}f_{j}e^{-i \vec{Q}\cdot(X\vec{r}_j + m_1X\vec{a} + m_2X\vec{b})} \\ = 
    F_{\text{unit}}^X \sum_{(m_1 , m_2) \in \mathcal{D}} e^{-i(m_1\vec{Q}\cdot (X\vec{a}) + m_2\vec{Q}\cdot (X\vec{b}))}
\end{align}
where $F_{\text{unit}}^X$ is the unit cell of a rotated layer with expression
\begin{align}
    F_{\text{unit}}^X = \sum_{j=1}^{N}f_{j}e^{-i \vec{Q}\cdot (X \vec{r}_j)}.
\end{align}
It follows that the structure factor of a rotated layer that is also rectangular is
\begin{align}
F^X = F_{\text{unit}}\frac{\sin(N_a \pi h_{X})}{\sin(\pi h_{X})}\frac{\sin(N_b \pi k_{X})}{\sin(\pi k_{X})}e^{-i(N_a-1)\pi h_{X}}e^{-i(N_b-1)\pi k_{X}}
\label{rotated_layer}
\end{align}
where 
\begin{align}
    h_{X} = \vec{Q} \cdot ( X \vec{a} ) \\
    k_{X} = \vec{Q} \cdot ( X \vec{b} ) \nonumber \\
    l_{X} = \vec{Q} \cdot ( X \vec{c} ). \nonumber
\end{align}
The next example is a layer that differs only by a translation $\vec{v}$ from a layer with structure factor $F$. The translated layer has structure factor $F^{\vec{v}}$ related to $F$ via the simple relation
\begin{align}
    F^{\vec{v}} = Fe^{-i \vec{Q}\cdot\vec{v}}.
\end{align}
Usefully, when layers differ by some linear transformation (rotation or translation) the structure factor of each layer can be expressed as a periodic arrangement of unit cells with translational symmetry, where each layer type's unit cell is related by some transformation to the unit cell of another layer type. However, this feature does not apply to layers that differ by some nonlinear transformation like change in curvature, which was discussed by \citeasnoun{LI20071686} when describing disordered layers of graphite. In fact for a reference layer with structure factor $F$, a second layer differing from the reference by a nonlinear transformation $\phi$ has structure factor
\begin{align}
    F^{\phi} = \sum_{(m_1 , m_2) \in \mathcal{D}}\sum_{j=1}^{N}f_{j}e^{-i \vec{Q}\cdot\big(\phi(\vec{r}_j + m_1\vec{a} + m_2\vec{b})\big)}.
\end{align}
The nonlinearity of $\phi$ means we cannot factorise out the structure factor of a unit cell; which is consistent with physical intuition. One would not expect a curved layer to comprise a periodic array of identical unit cells because some cells would be curved more than others. Consequently, instead of thinking about aperiodic crystals as comprised of unit cells, it is safer to think of them as comprised of layers, that cannot (in general) be broken down into constituent cells.

With the preamble about structure factors out the way, we are in a position to approach the differential scattering cross section (or scattering pattern) of an aperiodic crystal. First of all, suppose a crystal is composed of a sequence of layers, labelled in order from $n = 0, ... ,N_c$. Each layer has a type (or structure factor) indexed by the set $\mathcal{A}$. If the number of layer types is finite, then $\mathcal{A}$ is some finite subset of the positive integers $\mathbb{N}$. If $\mathcal{A}$ is countably infinite then we let $\mathcal{A} = \mathbb{N}$ and if uncountably infinite we allow $\mathcal{A} \subset \mathbb{R}^n$ to be open and connected. The structure factor of each layer is labelled $F_n$, so the structure factor of the entire crystal $\psi$ is
\begin{align}
    \psi = \sum_{n=0}^{N_c} F_n e^{2 \pi i n l}
\end{align}
and it follows that the cross section has expression
\begin{align}
    \frac{d \sigma}{d \Omega} = \abs{\psi}^2 =
    \sum_{n=0}^{N_c}\sum_{m=0}^{N_c} F_n F^*_{m} e^{2 \pi i (n - m) l}.
\end{align}
Now, $F_n F_m^*$ is the average structure factor product $Y_{m-n}$ discussed by \citeasnoun{PhysRevB.34.3586}, obtained by taking expectation over the distribution of structure factor pairs separated by $m-n$ layers. When $\mathcal{A}$ is countable, we can write this down as
\begin{align}
    Y_{m-n} = F_n F_m^* = \sum_{x \in \mathcal{A}}\sum_{y \in \mathcal{A}}F(x)G_{m-n}(x,y)F^*(y)
\end{align}
where $G_{m}(x,y)$ is the pair correlation function between layers $x,y \in \mathcal{A}$ where $y$ is $m$ layers ahead of $x$, and $F(x)$ and $F(y)$ are the structure factors of $x$ and $y$ respectively. If $\mathcal{A} \subset \mathbb{R}^n$ then similarly
\begin{align}
        Y_{m-n} = \int_{ \mathcal{A}}\int_{ \mathcal{A}}F(x)G_{m-n}(x,y) F^*(y) dx dy.
        \label{Average_structure_factor_product}
\end{align}
    We can then recover the expression for the differential scattering cross section presented by \citeasnoun{PhysRevB.34.3586} and derived by \citeasnoun{wilson1942imperfections}
\begin{align}
    \frac{d \sigma}{d \Omega} = 
    \sum_{n=0}^{N_c}\sum_{m=0}^{N_c} Y_{m-n} e^{2 \pi i (n - m) l} \\ =
    \sum_{m_3 =-N_c}^{N _c}(N_c - \abs{m_3})Y_{m_3} e^{2 \pi i m_3 l}.
    \label{Cross_section}
\end{align}
For completeness, we note that the dimension of the crystal $N_a, N_b, N_c$ may not be the same for every crystal in a sample, but could in general follow some distribution. \citeasnoun{UNGAR2002929} for example, report that carbon blacks have log-normal size distribution. In this case the observed scattering pattern would be obtained by  summing the cross section over each dimension times the probability of a crystallite adopting that dimension.

\subsection{Powder averaging}

\begin{figure}
\includegraphics[]{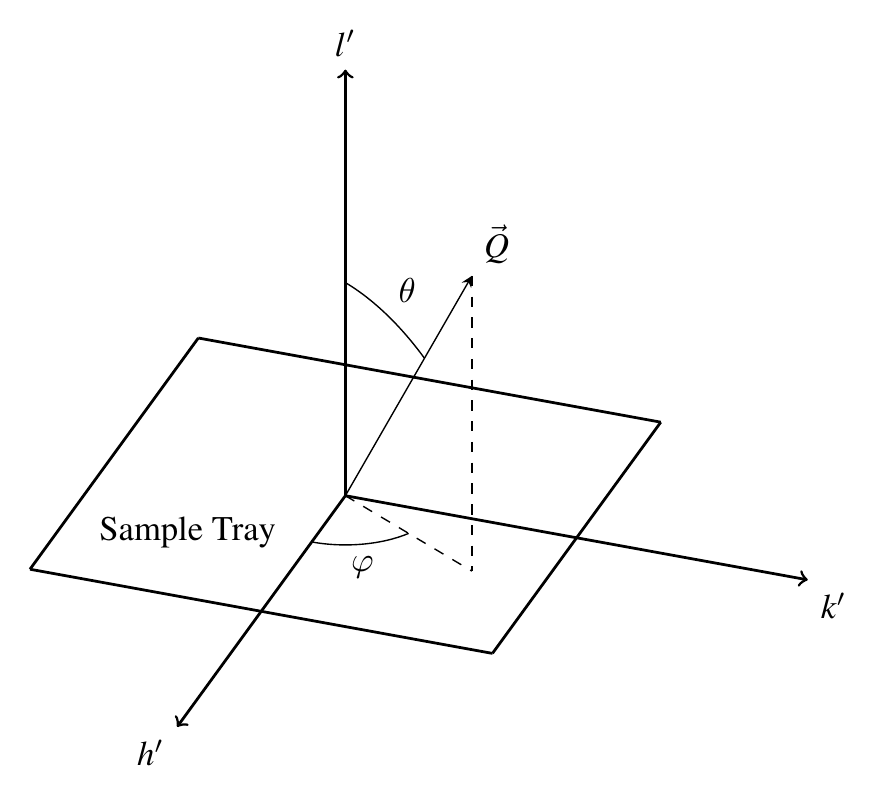}
\label{polar_coords}
\caption{The probability density that a crystallite centred at the origin is oriented such that it's normal vector is in the direction of the vector $\vec{Q}$ with polar angle $(\theta,\varphi)$ is $\omega(\vec{Q})$. A point in the reciprocal lattice coordinates $\vec{Q} = (h,k,l)$ is represented in Cartesian coordinates by $(h',k',l')$. The vector normal to the sample tray $\hat{n}$ is parallel to $l'$.}
\end{figure}

Suppose a powder sample of a crystal is placed into a flat tray with normal vector $\hat{n}$. The powder average $I(Q)$ of the cross section is given by
\begin{align}
    I(Q) = \int_{\partial B_Q} \frac{d \sigma}{d \Omega} (\vec{Q}) \omega(\vec{Q}) \ dS(\vec{Q})
\end{align}
where we are integrating over the sphere of radius $Q = |\vec{Q}|$ and have introduced the spherically symmetric preferred orientation function $\omega(\vec{Q})$ to represent the probability density that crystallite's normal vector is rotated by angles $(\theta, \varphi)$ from $\hat{n}$ where $(\theta, \varphi)$ are the spherical polar angles of the vector $\vec{Q}$. An illustration of this is shown in Figure \ref{polar_coords}. 
The preferred orientation function $\omega$ is introduced because crystallites in a container often align with the geometry of the container, resulting in some orientations being more likely than others. In the special case that all orientations are equally likely,
\begin{align}
    I(Q) = \frac{1}{4 \pi Q^2}\int_{\partial B_Q} \frac{d \sigma}{d \Omega} (\vec{Q}) \ dS(\vec{Q}).
\end{align}
Numerically computing either of these integrals is not easy because the the cross section $\frac{d \sigma}{d \Omega}(\vec{Q})$ is, roughly speaking, close to zero everywhere except for points surrounding $\vec{Q}$ where $h,k,l$ are all integers. At these points the cross section is highly peaked. As the size of a crystallite grows the peaks become taller and thinner, converging to delta functions  in the limit of infinite crystallite size. Na\"{i}ve quadrature does not perform well on integrands with many thin peaks, so should be avoided for computing the powder average of big crystallites. If the crystals are indeed big, a common method of computing the powder average is to numerically integrate each peak separately and sum the contributions. One can also employ the tangent-cylinder approximation derived by \citeasnoun{doi:10.1107/S0365110X51001409} and discussed by \citeasnoun{Shihw0018} to speed up the integration of each peak.

An alternative to numerical integration is to derive an expression for the powder average using a Harmonic expansion, which does not require numerical integration over the sphere! We shall present a version of this for the simplest case that $\omega(\vec{Q}) = 1 / 4\pi Q^2$, which may be adequate for a highly disordered material. Let $L^{m_3}_{ij}(x,y)$ denote the distance between the $i$th atom in a layer of type $x$ and the $j$th atom of a layer type $y$ for layers $x,y$ separated vertically by $m_3$ layers. When we talk about distance, we assume a unit length is $2 \pi |\vec{c}^*|$.
It is shown in Appendix \ref{powder_average_appendix} that
\begin{align}
    I(Q) =
    \sum_{m_3 =-N_c}^{N _c}(N_c - \abs{m_3}) \int_{\mathcal{A}}\int_{\mathcal{A}} G_{m_3}(x,y) \\ \times \sum_{i = 1}^{n_x} \sum_{j = 1}^{n_y} f_i f_j \text{sinc}\big(QL^{m_3}_{ij}(x,y)\big) dx dy. \nonumber
    \label{powder_avg}
\end{align}
The term
\begin{align}
\sum_{i = 1}^{n_x} \sum_{j = 1}^{n_y} f_i f_j\text{sinc}\big(QL^{m_3}_{ij}(x,y)\big)
\end{align}
is highly related to Debye's equation, who's 100th birthday was recently celebrated by \citeasnoun{Scardi:me0628}. 

\section{Finitely many hidden states}

It remains now to define the pair correlation function $G_m(x,y)$ which captures the probability of sampling from a crystal a layer of type $x$, then finding a layer of type $y$ $m$ layers ahead of $x$. To this end, we maintain the assumption of \citeasnoun{varn2013machine}, \citeasnoun{riechers2015pairwise},  and \citeasnoun{Varn201547} that the sequence of layers follows a Hidden Markov Model. In particular when the set of hidden states $\mathbb{S}$ and layer types $\mathcal{A}$ are finite, the Hidden Markov Model (HMM) is an ordered quintuple $\Gamma = (\mathcal{A},\mathbb{S},\mu_0,\mathcal{T},V)$ where the terms are exactly those defined by \citeasnoun{MarkovPaper1} in their Appendix A. 

In particular the probability of a layer adopting a hidden state $j \in \mathbb{S}$ can be represented as the element of a vector $v$. Given the hidden state of the HMM is $i \in \mathbb{S}$, then the probability of a transition to $j \in \mathbb{S}$ is the $ij$th element of a transition matrix $\mathcal{T}$. This matrix represents an operator which maps a distribution of hidden states $v$ of some layer to the distribution of hidden states $w$ of the next layer, which is to say 
\begin{align}
    \mathcal{T} v = w.
\end{align}
For a layer with hidden states following a distribution $v$, the layer found $m$ layers ahead has hidden state following the distribution $u$ which is related to $v$ by
\begin{align}
    \mathcal{T}^m v = u.
\end{align}
The stationary distribution $\pi$ of $\mathcal{T}$ represents the probability distribution over the set of hidden states obtained by sampling a layer from the crystal. A sufficient condition for $\pi$ to exist and be unique is that the Markov Chain induced by $\mathcal{T}$ is positive recurrent, which means from any state $s$ the probability of eventual return to $s$ state is unity.  Further since,
\begin{align}
    \mathcal{T} \pi = \pi
\end{align}
we have that $\pi$ is an eigenvector of $\mathcal{T}$ with eigenvalue $1$. 

Every hidden state emits a symbol from the alphabet according to some distribution that depends on the hidden state.
Even if the number of hidden states is finite, the alphabet of symbols $\mathcal{A}$ could be finite, countably infinite or uncountably infinite. The theory presented by \citeasnoun{riechers2015pairwise} and \citeasnoun{MarkovPaper1} assumes $\mathcal{A}$ is finite, and therefore that the probability distribution over symbols from the hidden state $s \in \mathbb{S}$ is a vector $v_s \in V$. Further, the probability of emitting a symbol $x \in \mathcal{A}$ is one of the entries of the vector $v_s$, denoted $v_s(x)$. This present paper extends the existing theory by stating that if $\mathcal{A}$ is countably infinite then $v_s$ is an infinite sequence with $x$th term $v_s(x)$ and if $\mathcal{A} \subset \mathbb{R}^n$ is uncountably infinite, then $v_s$ is a probability density function $v_s(x)$. To distinguish these cases, the ordered quintuple $\Gamma$ defining the HMM either contains vectors, sequences, or probability density functions for $\mathcal{A}$ finite, countably infinite and uncountably infinite respectively. Whatever the cardinality of $\mathcal{A}$, the pair correlation function $G_m(x,y)$ is given by
\begin{align}
    G_{m}(x,y) = \sum_{r \in \mathbb{S}}\sum_{s \in \mathbb{S}} v_r(x) \pi_{r}\mathcal{T}^m_{rs} v_s(y)
    \label{Pair_Correlation_Finite_S}
\end{align}
where $\mathcal{T}^m_{rs}$ is the $rs$th element of the matrix $\mathcal{T}^m$. With the expression for the pair correlation \eqref{Pair_Correlation_Finite_S} and cross section \eqref{Cross_section} together, we obtain a direct expression for the cross section of a crystal with finitely many hidden states, and any of finitely, countably infinitely or uncountably infinitely layer types. Section \ref{finite_state_space_uncountable_alphabet} runs through an application of this expression.

\subsection{A finite state space and uncountable alphabet}
\label{finite_state_space_uncountable_alphabet}

Suppose we have a finite state space and uncountable alphabet. Then for each state $r \in \mathbb{S}$ there is a probability density function $v_r(x)$ over the alphabet of symbols  $\mathcal{A} \subset \mathbb{R}^n$. It is shown in Appendix \ref{FSSUA} that the cross section for such a crystal can be expressed
\begin{align}
    \frac{d \sigma}{d \Omega} = Re \bigg\{\Tr \big( \Diag(\pi)H(2S + N_c I) \big) \bigg\} \label{Finite_state_space_cross_section}
\end{align}
where $H$ is a Hermitian matrix with dimension equal to that of $\mathcal{T}$ with $rs$th element
\begin{align}
    h_{rs} = \int_{\mathcal{A}} v_r(x)F(x) dx \int_{\mathcal{A}} v_s(y)F^*(y) dy.
\end{align}
Moreover
\begin{align}
    S = \sum^{N_c}_{m = 1}(N_c - m) (\mathcal{T}e^{2 \pi i l})^m
\end{align}
while $\Diag(\pi)$ is the diagonal matrix with elements the stationary vector $\pi$, $I$ is the identity matrix, $\Tr$ is the trace operator and $Re\{ z \}$ denotes the real part $z \in \mathbb{C}$.
It may be useful to note that
\begin{align}
    S = \sum^{N_c}_{m = 1}(N_c - m) (\mathcal{T}e^{2 \pi i l})^m \\ =
    \mathcal{T} e^{2 \pi i l} \big( (\mathcal{T} e^{2 \pi i l})^{N_c} - N_c( \mathcal{T} e^{2 \pi i l} - I ) - I \big)\big( \mathcal{T} e^{2 \pi i l} - I \big)^{-2}
\end{align}
when $\big( \mathcal{T} e^{2 \pi i l} - I \big)^{-2}$ exists.
This model includes a class of crystals that, in the absence of turbostratic disorder, comprise a finite number of layer types, where the probability of some layer type following another depends on the previous layer type. Each hidden state represents a layer type without turbostratic disorder, while the distribution over the alphabet of symbols represents the distribution over possible disorders a particular layer type could adopt. A simple, if perhaps unrealistic, example is a crystal composed of $2$ layer types labelled $A$ and $B$, which adopt some turbostratic disorder like a rotation, translation, or nonlinear deformation over some distributions $v_1(x)$ and $v_2(x)$ respectively. Suppose the probability given a layer is type $A$ that the next is also type $A$ is $\alpha$ and the probability that if a layer is type $B$ that the next will be type $A$ is $\beta$. Then the transition matrix between hidden states $A$ and $B$ takes the form
\begin{align}
    \mathcal{T} = 
    \begin{bmatrix}
    \alpha       & 1-\alpha \\
    \beta     & 1-\beta 
\end{bmatrix}.
\end{align}
For this toy model, we can expand the expression for the cross section $\eqref{Finite_state_space_cross_section}$ and arrive at
\begin{align}
    \frac{d \sigma}{d \Omega} = \frac{2}{(1-\alpha)(\beta - \alpha + 1)} \\ \times
    Re \bigg\{ s_1\big[\beta(h_{11} - h_{12}) + (1-\alpha)(h_{22} - h_{21})\big] \nonumber \\ + 
    s_2\big[ (1-\alpha)(h_{11} + h_{12}) + \beta(h_{22} + h_{21}) \big]\bigg\} \nonumber \\ +
    \frac{h_{11} + h_{22}}{1 - \alpha} \nonumber
\end{align}
where
\begin{align}
s_{1} = 
\begin{cases}
\frac{N_c}{2}(N_c - 1) \text{ if $e^{2 \pi i l} = 1$ }  \\
\frac{e^{2 \pi i l} (e^{2 \pi i l N_c} + N_c(1 - e^{2 \pi i l} ) - 1)}{(1 - e^{2 \pi i l})^2} \text{ otherwise,}
\end{cases}
\end{align}
and
\begin{align}
s_{2} = 
\begin{cases}
\frac{N_c}{2}(N_c - 1) \text{ if $(\alpha - \beta)e^{2 \pi i l}$ = 1}  \\
\frac{(\alpha-\beta) e^{2 \pi i l} ( (\alpha-\beta)^{N_c} e^{2 \pi i l N_c} + N_c(1 -(\alpha-\beta) e^{2 \pi i l} ) - 1)}{(1 - (\alpha-\beta) e^{2 \pi i l})^2} \text{ otherwise.}
\end{cases}
\end{align}
We can see that multiplying out the matrices and taking the trace generates an expression that is long and hard to read even for the simplest case of a crystal with 2 hidden states! Consequently, we consider a cross section defined once we have determined the transition matrix $\mathcal{T}$, the stationary distribution $\pi$ and the matrix $H$.
We will now explore a more sophisticated model, with a concrete application to carbon blacks.

\subsubsection{Recovery of Shi's model}

 \citeasnoun{ShiThesis} wrote a thesis about the crystal structure of disordered carbons to better understand their role as an electrode in lithium-ion batteries. Part of this document includes two sophisticated models of turbostratic carbon blacks, which can be fitted to scattering data using the program CARBONX written by \citeasnoun{Shihw0018}. CARBONX was recently picked up by \citeasnoun{ZHOU201417} who compared the performance of Shi's model to the standard Rietveld refinement program GSAS for describing the cross section of disordered carbons obtained from a range of sources.
\citeasnoun{ZHOU201417} found that Shi's account of turbostratic disorder improved the fit, suggesting the turbostratic disorder is much like \citeasnoun{ShiThesis} describes.

The remainder of this section will express both Shi's 1 layer model and 2 layer model as hidden Markov models, where each hidden states emits a disordered layer over some distribution dependent on the state. For both of these models, we will obtain the transition matrix $\mathcal{T}$, the matrix $H$ and stationary vector $\pi$, hence arrive at an expression for the cross section. We'll start with the 1 layer model, noting that these carbon blacks have 4 hidden states, we will label 1, 2, 3, 4. States $1,2,3$ enumerate the layer types $A,B,C$ while the hidden state $4$ represents a layer that has slipped across the basal plane in a random direction with random magnitude with uniform probability. According to Shi's 1 layer model model, for some probability $P$ of slippage across the basal plane, the transition matrix looks like
\begin{align}
    \mathcal{T} =
\begin{bmatrix}
    0       & \frac{1 - P}{2} &\frac{1 - P}{2} &P \\
     \frac{1 - P}{2}     & 0 & \frac{1 - P}{2} & P \\
     \frac{1 - P}{2}     & \frac{1 - P}{2} & 0 & P \\
     \frac{1 - P}{3}     & \frac{1 - P}{3} & \frac{1 - P}{3} & P
\end{bmatrix}
\end{align}
which has stationary vector
\begin{align}
    \pi = \frac{1}{4}
    \begin{bmatrix}
     1 \\
     1 \\
     1 \\
     1
\end{bmatrix}.
\end{align}
In addition to the possibility of a layer slipping across the basal plane, Shi's model stipulates that all layers may be shifted in the direction orthogonal to the basal plane. The probability of no shift occurring is denoted $g$, but if some shift does occur, the shift adopts a magnitude following a normal distribution centred at zero with variance $\sigma^2$. The probability density of a layer $n = 1, 2, 3$ being displaced by $z$ in the direction orthogonal to the basal plane therefore has expression
\begin{align}
    w(z) = g \delta(z) + (1-g)\frac{1}{\sqrt{2 \pi}\sigma}\exp\bigg(\frac{-z^2}{2 \sigma^2}\bigg). \label{norm_dist}
\end{align}
The alphabet $\mathcal{A}$ for Shi's 1 layer model is uncountable and comprises ordered pairs $x = (z,n)$ where $n \in \{1,2,3,4 \}$ denotes whether the 0 disorder layer is type $A,B,C$ or the 4th type that slipped across the basal plane, while $z$ is the displacement of that layer orthogonal to the basal plane and follows distribution $\eqref{norm_dist}$. 

The structure factor of a layer $x \equiv (n,z)$ can therefore be written
\begin{align}
    F(x) = F(z,n) = F_n(z) 
\end{align}
and we have that a layer $A$ (which has hidden state $1$) with 0 displacement orthogonal to the basal plane has unit cells with a structure factor
\begin{align}
    F_1^{\text{unit}}(0) = 2 f \cos\bigg( \frac{2 \pi}{3}(h+k) \bigg)
\end{align}
where $f$ is the form factor of a carbon atom. Consequently, if we make the simplifying assumption that all layers are rectangular with the same dimensions, then the structure factor of the layer $A$ with 0 orthogonal displacement is (by equation \eqref{structure_factor_of_layer})
\begin{align}
    F_1(0) = 2 f \cos\bigg( \frac{2 \pi}{3}(h+k) \bigg)\frac{\sin(N_a \pi h)}{\sin(\pi h)}\frac{\sin(N_b \pi k)}{\sin(\pi k)} \\ \times e^{-i(N_a-1)\pi h}e^{-i(N_b-1)\pi k}. \nonumber
\end{align}
The layers $A,B,C$ with displacement $z$ have structure factors
\begin{align}
    &F_1(z) = F_1(0) e^{2 \pi i z l} \\
    &F_2(z) = F_1(0) e^{2 \pi i( z l + (h + k)/3)} \nonumber \\
    &F_3(z) = F_1(0) e^{2 \pi i( z l - (h + k)/3)} \nonumber
\end{align}
respectively. Now the probability of a hidden state $n$ emitting a symbol $x \in \mathcal{A}$ is given by the probability density function $v_n(x)$ so
\begin{align}
    \int_{\mathcal{A}}v_n(x) F(x) dx =  \int_{\mathbb{R}}w(z) F_n(z) dz \\ =
    e^{2 \pi i \phi_n}F_1(0)\int_{\mathbb{R}}w(z) e^{2 \pi i zl} dz \\=
    e^{2 \pi i \phi_n}F_{1}(0)\mathcal{F}[w](l)
\end{align}
where $\mathcal{F}[w]$ is the Fourier transform of $w$
\begin{align}
    \mathcal{F}[w](l) = g + (1-g)\exp \bigg( -\frac{ \sigma^2  l^2}{2} \bigg)
\end{align}
and
\begin{align}
    \phi_n =
    \begin{cases}
    0 &\text{ if $n = 1$} \\
    (h+k)/3 &\text{ if $n = 2$} \\
    -(h+k)/3 &\text{ if $n = 3$}
    \end{cases}
\end{align}
is introduced for notational convenience.
Shi made the assumption that total contribution to the scattering pattern from the layers translated across the basal plane is zero, so we choose $v_4(x)$ and $F(x)$ such that
\begin{align}
    \int_\mathcal{A} v_{4}(x)F(x)dx = 0.
\end{align}
This gives us an expression for $H$
\begin{align}
    H = \abs{F_1(0)\mathcal{F}[w](l)}^2 \\
    \times
    \begin{bmatrix}
    1       & e^{-2/3 \pi i(h + k)} &e^{2/3 \pi i(h + k)} &0 \\
     e^{2/3 \pi i(h + k)}     & 1 & e^{-2/3 \pi i(h + k)} & 0 \\
     e^{-2/3 \pi i(h + k)}     & e^{2/3 \pi i(h + k)} & 1 & 0 \\
     0     & 0 & 0 & 0
\end{bmatrix} \nonumber.
\end{align}
With $\mathcal{T}$, $\pi$ and $H$ we have all we need to evaluate equation \eqref{Finite_state_space_cross_section} and obtain the cross section for Shi's 1 layer model.

Shi's 2 layer model is similar, and in the formalism of this paper has 7 hidden states each comprising pairs of conventional layers $AB$, $AC$, $BA$, $BC$, $CA$, $CB$ as well as a layer $XX$ translated somewhere across the basal plane. Like the 1 layer model, layers are displaced in the direction orthogonal to the basal plane according to distribution \eqref{norm_dist}, but this time with $g = 0$. We enumerate these layer types from $1$ to $7$ and obtain the transition matrix according to Shi's description
\begin{align}
\mathcal{T} = 
    \begin{bmatrix}
    P_t       & 0 & 0 & 0 & \bar{P}  & 0 & P \\
     0     & P_t & \bar{P}  & 0 & 0 & 0 & P \\
     0     & 0 & P_t & 0 & 0 & \bar{P}  & P \\
     \bar{P}     & 0 & 0 & P_t & 0 & 0 & P \\
     0     & 0 & 0 & \bar{P}  & P_t & 0 & P \\
     0     & \bar{P}  & 0 & 0 & 0 & P_t & P \\
     \frac{1-P}{6}     & \frac{1-P}{6} & \frac{1-P}{6} & \frac{1-P}{6} & \frac{1-P}{6} & \frac{1-P}{6} & P
\end{bmatrix}
\end{align}
where $\bar{P} = 1 - P_t - P$ and $P_t$, $P$ and $\bar{P}$ are probabilities summing to 1. The stationary vector is
\begin{align}
    \pi = \frac{1}{7}
    \begin{bmatrix}
     1 \\
     1 \\
     1 \\
     1 \\
     1 \\
     1 \\
     1
    \end{bmatrix}.
\end{align}
Now to obtain $H$, first let
\begin{align}
    &\varphi_1 = 1 + e^{2 \pi i( c l + (h + k)/3)} \\
    &\varphi_2 = 1 + e^{2 \pi i( c l - (h + k)/3)} \nonumber \\
    &\varphi_3 = e^{2 \pi i(h + k)/3} + e^{2 \pi i c l} \nonumber \\
    &\varphi_4 = e^{2 \pi i (h + k)/3} + e^{2 \pi i( c l - (h + k)/3)} \nonumber \\
    &\varphi_5 = e^{- 2 \pi i (h + k)/3} + e^{2 \pi i c l} \nonumber \\
   &\varphi_6 = e^{ - 2 \pi i (h + k)/3} + e^{2 \pi i ( c l + (h + k)/3)} \nonumber \\
   &\varphi_7 = 0, \nonumber
\end{align}
which we have introduced for notational convenience.
The structure factor for layer types $(n,z)$ are 
\begin{align}
    F_n(z) = F_1(0)e^{2 \pi i z l}\varphi_n
\end{align}
for $n = 1$ to $6$. Now the matrix $H$ has elements
\begin{align}
    h_{nm} = \abs{F_1(0)\mathcal{F}[w](l)}^2\varphi_n\varphi_m,
\end{align}
where we have for $n > 6$ or $m > 6$ that $h_{nm} = 0$ by construction, because the 2 layer model (similar to the 1 layer model) assumes
\begin{align}
    \int_\mathcal{A} v_{7}(x)F(x)dx = 0.
\end{align}
Both of Shi's models make specific assumptions that simplify the mathematics and allow the models to be expressed concisely, but are not necessarily physically principled. For example the 2 layer model accounts for normally distributed turbostratic spacing between pairs of layers, but not for disorder within a pair of layers. Moreover, certain transitions e.g. $AB$ to $AC$ are assumed impossible, even though they are physically plausible. By framing Shi's model in the HMM framework, we can straight forwardly modify the model to encompass any disorder we like, while retaining a neat expression for the cross section. Recommending specific improvements to Shi's model is beyond the scope of this paper, which instead presents these examples to demonstrate that the HMM framework is flexible and general enough to describe a wide range of turbostatic materials.

\section{Uncountably many hidden states}

Having examined a HMM with a finite number of hidden states, we will now move on the stranger world of uncountably many hidden states.
If $\mathbb{S}$ is uncountably infinite then a probability distribution over $\mathbb{S}$ is given by some probability density function $v$. We suppose $\mathbb{S} \subset \mathbb{R}^n$ is open, connected and bounded. Since the integral of $v$ over $\mathbb{S}$ must equal unity, $v$ is necessarily square integrable and therefore in the Hilbert space of square integrable functions $L^2$. Given the states are distributed according to $v$, the distribution over hidden states at the next layer $w \in L^2$ is
\begin{align}
    \int_{\mathbb{S}}k(r,s)v(r) dx = w(s)
\end{align}
where $k(r,s)$ represents the probability density of $s \in \mathbb{S}$ following $r \in \mathbb{S}$ and is called the transition kernel. This gives rise to an integral operator $\mathcal{T}:L^2 \to L^2$ defined
\begin{align}
    (\mathcal{T}v)(s) = \int_{\mathbb{S}}k(r,s)v(r) dx.
\end{align}
The probability of sampling from the crystal a layer with hidden type $r$ is given by the probability density function $\pi(r)$ which exists, is unique and satisfies
\begin{align}
    \mathcal{T}\pi = \pi
\end{align}
 if the transition kernel $k(r,s)$ is positive recurrent.
If it exists, the stationary distribution $\pi$ is an eigenvector of the operator $\mathcal{T}$ with eigenvalue one.
Given a distribution over hidden states $v$, the distribution over hidden states of a layer $w$ after $m$ transitions satisfies
\begin{align}
    \mathcal{T}^m v = w.
\end{align}
The pair correlation function for a crystal with uncountably many hidden states is therefore
\begin{align}
     G_m(x,y) = \int_{\mathbb{S}}\int_{\mathbb{S}} v(r,x)\pi(r)(\mathcal{T}^m \delta_r)(s)v(s,y) dr ds
\end{align}
where
$\delta_r(s)$ is the shifted delta function $\delta(r-s)$ where we interpret
\begin{align}
    (\mathcal{T}^m\delta_r)(s) = (\mathcal{T}^{m - 1}\mathcal{T}\delta_r)(s) \\
    = (\mathcal{T}^{m - 1}k_r)(s)
\end{align}
where $k_r$ is the probability density function $k_r(s) \equiv k(r,s)$. 

\subsection{Special case of a Markov chain}

Suppose the probability of a layer being a certain type depends only on the type of the previous layer, then we have a Markov chain of layer types. This is a special case of a HMM where every hidden state emits a symbol with probability 1 and no two states emit the same symbol. Formally, this is obtained by letting $\mathbb{S}=\mathcal{A}$ and letting $V$ be the identity map. For a Markov chain of layers adopting one of uncountably many layer types, the pair correlation function reduces to
\begin{align}
    G_m(x,y) =  \pi(x)(\mathcal{T}^m \delta_x)(y).
\end{align}
With this, we show in Appendix \ref{UncountableAppendix} that the cross section of a crystal described by a Markov chain, with an uncountable infinite of layer types is
\begin{align}
\frac{d \sigma}{d \Omega} = \label{UCCS}
2 Re \Bigg\{ \int_{\mathcal{A}}\int_{\mathcal{A}}F(x)F^*(y)\pi(x)Z\delta_x(y) dx dy \Bigg\}  \\
+ N_c \int_{\mathcal{A}} \abs{F(x)}^2 \pi(x) dx \nonumber
\end{align}
where $Re\{z\}$ represents the real part of the complex number $z \in \mathbb{C}$, while $\delta_x(y)$ is the shifted delta function $\delta(x-y)$ and $Z:L^2 \to L^2$ is a linear operator defined
\begin{align}
Z &\equiv \sum_{m_{3} = 1}^{N_c}(N_c - \abs{m_3})(e^{2 \pi i l}\mathcal{T})^{m_3}
\end{align}
where we interpret $Z\delta_x(y)$ as the evaluation at $y$ of the function $Z\delta_x$.

Expression \eqref{UCCS} for the cross section is quite unwieldy, demanding the evaluation of both a double integral and repeated application of the operator $\mathcal{T}$. Using numerical integration for this task may not be a good idea. A possible approach is to approximate the infinite state space as large but finite, hence discretising the structure factors and state distributions - collapsing the problem to the case of a large but finite state space.

Alternatively, one can follow the lead of \citeasnoun{PhysRevB.34.3586}, \citeasnoun{Hansen2008}, \citeasnoun{0953-8984-20-28-285105} and compute the cross section using a Monte Carlo simulation. When the state space $\mathbb{S}$ is finite, \citeasnoun{MarkovPaper1} argue that the Monte Carlo approach is much slower than computing the cross section explicitly using matrix operations. However, for an uncountable state space the problem of computing the cross section explicitly boils down to recursively computing integrals, which is generally much harder.  
In the upcoming sections, we consider special cases of \eqref{UCCS} that admit to further analysis and yield expressions that are faster to compute and perhaps more informative.
\subsection{Compact and self-adjoint transition operator}
The first of these cases requires that $\mathcal{T}$ is a compact, self-adjoint operator on the Hilbert space of square integrable functions $L^2$. These conditions hold for a crystal where the probability density of a state $y$ following state $x$ is equal to the probability density of state $x$ following $y$; which is to say $k(x,y) = k(y,x)$. This condition is sufficient (but not necessary) to imply that the Markov chain of layers describing the crystal is \emph{reversible}, representing a special type of crystal that appears the same (in some statistical sense) when turned upside-down. The significance of these reversible crystals is treated extensively by \citeasnoun{Ellison2009} and discussed in the context of ice and opal by \citeasnoun{MarkovPaper1}. For a crystal comprising layers that differ (for example) only by some rotation or translation, the probability density of a layer rotated to angle or position $y$ following a layer rotated to angle or position $x$ must equal the probability density of a layer at angle or position $x$ following one at angle or position $y$. We note here that this idea could in principle apply to a much broader class of disorders.

With the assumption that $\mathcal{T}$ is a compact self-adjoint operator on some Hilbert space, the Spectral Theorem provides an expression for the cross section
\begin{align}
\frac{d \sigma}{d \Omega} = \label{UCCSadj}
2 Re \Bigg\{ \sum_{n \in \Lambda} s_n\int_{\mathcal{A}}F(x) \pi(x) u_n(x) dx \int_{\mathcal{A}} F^*(y) u_n(y)  dy \Bigg\} \\
+ N_c \int_{\mathcal{A}} \pi(x) \abs{F(x)}^2 dx \nonumber
\end{align}
where
\begin{align}
s_{n} = 
\begin{cases}
\frac{N_c}{2}(N_c - 1) \text{ if $\lambda_n e^{2 \pi i l}=1$ }  \\
\frac{\lambda_n e^{2 \pi i l} ( \lambda_n^{N_c} e^{2 \pi i l N_c} + N_c(1 - \lambda_n e^{2 \pi i l} ) - 1)}{(1 - \lambda_n e^{2 \pi i l})^2} \text{ otherwise,}
\end{cases}
\label{s_n}
\end{align}
and $u_n$ and $\lambda_n e^{2 \pi i l}$ the eigenvectors and eigenvalues of $e^{2 \pi i l}\mathcal{T}$ indexed by the set $\Lambda$ which repeats eigenvalues according to their algebraic multiplicity. The details are fleshed out in Appendix \ref{self_adj_app}.

\subsection{A convolution kernel} 
\label{conv_ker}

The second case applies to a so-called convolution kernel $k$ on an uncountable state space, which requires for some $n \in \mathbb{N}$ that $\mathcal{A}$ is the open hypercube of dimension $n$ denoted $\mathcal{A} = (0,1)^n$ and that
\begin{align}
k(x,y) \equiv \sum_{m \in \mathbb{Z}^n}P(m+y-x) 
\label{conv_ker_defn}
\end{align}
for some probabilty distribution $P$ in the Hilbert space of square integrable functions $L^2$. If we return to the example of a crystal composed of layers that differ only by a rotation, we can interpret condition \eqref{conv_ker_defn} as insisting that the angle of rotation between any pair of layers follows the same probability distribution $P$. The summation over $m$ represents the fact that a rotation to angle $\theta$ is equal to a rotation to angle $\theta + 2m\pi$ for all $m \in \mathbb{Z}$ so we let the space $\mathcal{A} = (0,1)$ and interpret for $x \in \mathcal{A}$ that $2\pi x$ is a layer's angle of rotation. Now $n$ represents the dimension of the state space $\mathcal{A}$, and equals 1 here. If for example, layers were identical up to some translation in any of three directions, then the state space $\mathcal{A}$ would be three dimensional and $n$ would adopt the value 3 and $\mathcal{A} = (0,1)^3$.

We note that $P(y-x) = P(x-y)$ does not hold in general, so $\mathcal{T}$ is not necessarily self-adjoint even if it has a convolution kernel. We also remark that the stationary distribution $\pi$ of the operator $\mathcal{T}$ with convolution kernel is uniform. With this established, we present in Appendix \ref{UncountableAppendix} a derivation for the cross section of a crystal with kernel $P$
\begin{align}
\frac{d \sigma}{d \Omega} =  
Re \Bigg\{ \int_{\mathcal{A}}2F(x)(F^* \circledast s)(x) + N_c \abs{F(x)}^2 dx \Bigg\} \label{UCCS_wS} 
\end{align}
with $a \circledast b$ representing the convolution of $a$ with $b$ and the function $s \in L^2$ satisfying
\begin{align}
\mathcal{F}[s] = 
\begin{cases}
\frac{N_c}{2}(N_c - 1) \text{ if $\mathcal{F}[P]e^{2 \pi i l} = 1$}  \\
\frac{\mathcal{F}[P] e^{2 \pi i l} ( \mathcal{F}[P]^{N_c} e^{2 \pi i l N_c} + N_c(1 - \mathcal{F}[P] e^{2 \pi i l} ) - 1)}{(1 - \mathcal{F}[P] e^{2 \pi i l})^2} \text{ otherwise,}
\end{cases}
\label{s_in_L2}
\end{align}
where $\mathcal{F}[\phi]$ represents the Fourier transform of $\phi \in L^2$. Given a choice of $P$, the function $s \in L^2$ does not have an analytic form in general, but can be approximated numerically using at most 2 Fast Fourier Transforms (FFTs). The first FFT is used to compute $\mathcal{F}[P]$, if the transform cannot be obtained analytically, from which we can find $\mathcal{F}[s]$ via equation \eqref{s_in_L2}. The second FFT is used to find the inverse transform of $\mathcal{F}[s]$, yielding $s$. That said, computing $s$ explicitly may not even be necessary if we observe that
\begin{align}
    F^* \circledast s = \mathcal{F}^{-1}\big[ \mathcal{F}[F^*]\mathcal{F}[s] \big]
\end{align}
by the convolution theorem.

\paragraph{Application to twisted $n$-layer graphene} 

With the theory outlined, we now have a lens through which to examine a toy model of twisted $n$-layer graphene. Suppose first of all that the layers of graphene can adopt any of the uncountably many angles of rotation $\theta \in [\frac{-\pi}{6},\frac{-\pi}{6}] = \mathcal{A}$ relative to some arbitrary $2$D coordinate system. We assume that the probability of a rotation to angle $y$ given a previous layer is at angle $x$ is given by a symmetric function $P \in L^2$ such that $P(y-x) \equiv P(x-y)$. Consequently, the cross section of this model satisfies both equations \eqref{UCCSadj} and \eqref{UCCS_wS}. In order to express the cross section more concretely, we first note that the structure factor of the graphene unit cell has expression

\begin{align}
    F^{\text{unit}}(0) = f e^{\frac{2}{3} \pi i (h + k) } + f e^{\frac{4}{3} \pi i (h + k) }
\end{align}
so the structure factor of the unit cell at some arbitrary rotation $\theta$ is therefore
\begin{align}
    F^{\text{unit}}(\theta) = f e^{\frac{2}{3} \pi i (h(\cos(\theta) - \sin(\theta)) + k(\cos(\theta) + \sin(\theta)))  } \\ +
    f e^{\frac{4}{3} \pi i (h(\cos(\theta) - \sin(\theta)) + k(\cos(\theta) + \sin(\theta)))}. \nonumber
\end{align}
Then by the derivation of equation \eqref{rotated_layer}, the structure factor of a graphene layer is
\begin{align}
    F(\theta) = F^{\text{unit}}(\theta)\frac{\sin(N_a \pi h_{\theta})}{\sin(\pi h_{\theta})}\frac{\sin(N_b \pi k_{\theta})}{\sin(\pi k_{\theta})} \\ \times e^{-i(N_a-1)\pi h_{\theta}}e^{-i(N_b-1)\pi k_{\theta}} \nonumber
\end{align}
where
\begin{align}
    h_{\theta} = \vec{Q} \cdot ( X \vec{a} ) \\
    k_{\theta} = \vec{Q} \cdot ( X \vec{b} ) \nonumber
\end{align}
where $X$ is the rotation matrix
\begin{align}
    X = 
    \begin{bmatrix}
    \cos(\theta)      & -\sin(\theta) \\
    \sin(\theta)     & \cos(\theta)
\end{bmatrix}.
\end{align}
Next, we observe that since $\mathcal{T}$ has a convolution kernel $P(y-x)$, the operator $\mathcal{T}$ has a uniform stationary distribution $\pi$, which integrates to unity over its domain $[\frac{-\pi}{6},\frac{\pi}{6}]$, so we deduce 
\begin{align}
    \pi(\theta) \equiv \frac{3}{\pi}.
\end{align}
Now, the angle of rotation between some pair of layers follows a distribution $P$, which would ideally be chosen with some physical motivation, and be consistent with empirical data, like that presented by \citeasnoun{doi:10.1021/nl204547v}. With a choice of $P$, we have an expression for the cross section of $n$-layer twisted graphene
\begin{align}
\frac{d \sigma}{d \Omega} =
\frac{6}{\pi} Re \Bigg\{ \int\displaylimits_{-\frac{\pi}{6}}^{\frac{\pi}{6}}F(\theta)(F^*\circledast s)(\theta) + N_c \abs{F(\theta)}^2 d\theta \Bigg\}
\end{align}
which can be numerically integrated in good time to high precision.

\subsection{A convolution kernel with layers identical up to translation}

\begin{figure}
  \caption{Here, the layers nearest the axis perpendicular to the basel plane have position normally distributed about their expected positions $0, c, 2c, 3c ...$ so exhibit disorder of the first type. The layers furthest from the axis have normally distributed pairwise separation hence undergo disorder of the second type and form a less coherent scattering pattern. The distributions have the same variance $0.1c$.} 
  \centering
\includegraphics[]{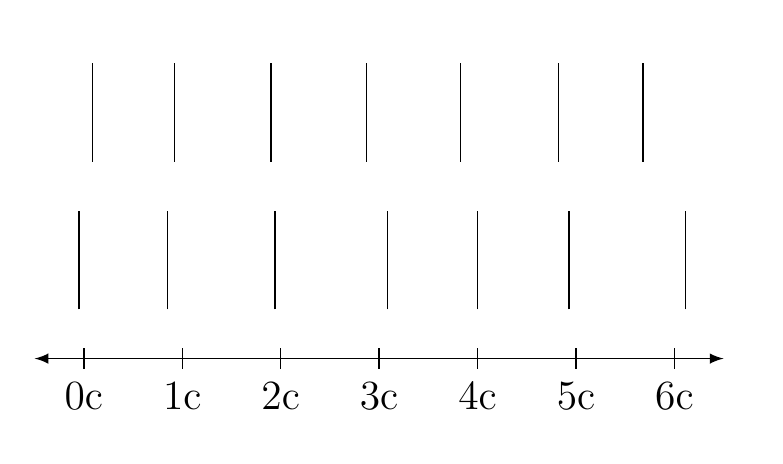}
\label{drift}
\end{figure}

We considered in section \ref{conv_ker} a crystal with transition operator $\mathcal{T}$ imbued with a convolution kernel where layers can exhibit a broad range of turbostratic disorder. In this section we zoom into a special case where all layers are identical up to some translation, and show in Appendix \ref{conv_ker_appendix} that the cross section of these crystals has expression
\begin{align}
\frac{d \sigma}{d \Omega} = \abs{F(0)}^2 \bigg( 2 Re\big\{ \mathcal{F}[s](\vec{Q}) \big\} + N_c\bigg).
\end{align}
One can intuit the relevance of this model by considering a crystal where a layer may slip across the basal plane by some magnitude following some distribution. For example the centre of one layer may slip by some magnitude away from the centre of the next layer. As we move up through the crystal, the centre of each layer performs a random walk, and the centre of the $n$th layer will gradually drift away from the centre of the 1st layer as $n$ grows. Alternatively, one might consider a sequence of layers with expected vertical separation $c$ (where vertical is orthogonal to the basal plane) but due to the effects of disorder, a layer is separated vertically from its predecessor by some random value following a normal distribution centred at $c$. \citeasnoun{Disorder_first_second} and \citeasnoun{Guinier1964} describe this type of disorder as \emph{disorder of the second type}. This is subtly different from Shi's model, where the layers adopt positions following independent and identical normal distribution centred at each of the layers' expected position, an example of \emph{disorder of the first type}. Figure \ref{drift} illustrates this difference.

This distinction is important because the different disorders would arise from different physics, and the different disorders give rise to different scattering patterns. In particular, Shi's model of disorder suggests that if a layer is separated by its neighbour by some distance approximately $c$, then the next layer is separated by approximately $2c$, and the next approximately $3c$, and this continues for arbitrary $nc$, without reduction in the accuracy of the approximation. However for the model incorperating disorder of the second type, this approximation would gradually get worse with increasing $n$. This suggests that the form of the scattering pattern, which depends strongly on the periodicity of layers, would differ, and this is reflected in the different expressions for the cross section.

To provide a specific example of disorder of the second type, suppose we have a sequence of graphite layers that are identical, except for some vertical displacement $z$ that follows a distribution
\begin{align}
    P(z) = g \delta(z-c) + (1-g) \frac{1}{\sqrt{2\pi}\sigma}\exp\bigg( \frac{(z-c)^2}{2 \sigma^2} \bigg) 
\end{align}
inspired by Shi's 1 layer model. Then by noting
\begin{align}
    |F(0)|^2 = 4f^2\cos \bigg(\frac{2 \pi}{3}(h+k)\bigg)^2 \frac{\sin(N_a \pi h)^2}{\sin(\pi h)^2}\frac{\sin(N_b \pi k)^2}{\sin(\pi k)^2}
\end{align}
we have all we need to compute the cross section explicitly. This model is overly simple of course, but admits to much extension, and could therefore capture a large range of possible disorders.

\section{A countable infinity of hidden states}

Having delved into both uncountable and finite state spaces, this section presents a short treatment of countably infinite state spaces. 
Suppose $\mathbb{S}$ is countably infinite, then the probability of a HMM adopting each state is enumerated as a sequence. Since this sequence sums to $1$ it is necessarily square summable hence an element of the Hilbert space of square summable sequences $\ell^2$. For a probability distribution $v \in \ell^2$ over hidden states, the probability distribution over states for the next state $w$ is given by
\begin{align}
    \sum_{i \in \mathbb{N}} k_{ij}v_j = w_i
\end{align}
where $k_{ij}$ is the transition kernel denoting the probability of the state $j$ following the state $i$.
Much like HMMs with finite and uncountable hidden states, the transition kernel gives rise to the transition operator $\mathcal{T} : \ell^2 \to \ell^2$ with stationary distribution an eigensequence with associated eigenvalue 1. 
The alphabet of symbols can be finite, countably infinite or uncountably infinite. In the special case that every state emits a unique symbol with probability 1, the HMM is just a Markov chain. This forms a simple model of $n$-layer moir\'{e} graphene, where the set of all layer pairs forming a moir\'{e} pattern is countably infinite. We call each of these layer pairs a superlattice and suppose each superlattice is labelled by some $i \in \mathbb{N}$ and given that a superlattice is type $i$ the probability that the next superlattice is type $j$ depends only on $i$ and $j$. Then the sequence of superlattices forms a Markov chain with uncountable state space $\mathbb{S}$. The cross section of a Markovian crystal with countably infinite state space is shown in Appendix \ref{UncountableAppendix} to satisfy
\begin{align}
\frac{d \sigma}{d \Omega} = \label{CCS}
\Bigg( 2 Re \Bigg\{ \sum_{i \in \mathbb{N}}\sum_{j \in \mathbb{N}}F_iF^*_j\pi_i Z\delta_{ij} \Bigg\}
+ N_c \sum_{i \in \mathbb{N}} F^*_i F_i \pi_i \Bigg) \nonumber
\end{align}
where $F_i$ is structure factor of the state indexed by $i$, $\pi_{i}$, is the $i$th element of the stationary distribution $\pi$,$\delta_{ij}$ is the Kronecker delta, and $Z: \ell^2 \to \ell^2$ is defined
\begin{align}
Z &\equiv \sum_{m_{3} = 1}^{N_c}(N_c - \abs{m_3})(e^{2 \pi i l}\mathcal{T})^{m_3}.
\end{align}
We note here that if $\pi$ is uniform, then for any sequence $\phi_i \in \ell^2$ we interpret
\begin{align}
\sum_{i \in \mathbb{N}} \pi_i \phi_i \equiv \lim_{n \to \infty}\frac{1}{n}\sum_{i = 0}^{n}\phi_i.
\end{align}
Moreover, if $k_{ij} = k_{ji}$ then by the Spectral Theorem
\begin{align}
\frac{d \sigma}{d \Omega} = \label{CCSadj}
2 Re \Bigg\{ \sum_{n \in \Lambda} s_n\sum_{i \in \mathbb{N}}F_i \pi_i u_n^i \sum_{j \in \mathbb{N}} F^*_j u_n^j  \Bigg\} 
+ N_c \sum_{i \in \mathbb{N}} \pi_i F^*_i F_i
\end{align}
with
\begin{align}
s_{n} = 
\begin{cases}
\frac{N_c}{2}(N_c - 1) \text{ if $\lambda_n e^{2 \pi i l}=1$ }  \\
\frac{\lambda_n e^{2 \pi i l} ( \lambda_n^{N_c} e^{2 \pi i l N_c} + N_c(1 - \lambda_n e^{2 \pi i l} ) - 1)}{(1 - \lambda_n e^{2 \pi i l})^2} \text{ otherwise,}
\end{cases}
\end{align}
where $u_n$ and $\lambda_n$ are the eigensequences and eigenvalues of $\mathcal{T}$ indexed by $\Lambda$ repeating eigenvalues according to their algebraic multiplicity. A derivation is presented in Appendix \ref{general_kernel}.

\section{Outlook}

We began to extend the study of chaotic crystallography to crystals with infinitely many layer types - describing turbostratically disordered materials like carbon blacks, smectites, and $n$-layer graphene. In particular, we derived an explicit cross section for carbon blacks related to the two models proposed by \citeasnoun{ShiThesis}.
There are many other disordered materials that could be examined under the framework presented here, smectites for example, are often studied with a qualitative look at diffraction peaks \cite{Ufer:2008:0009-8604:272} \cite{Turbostratic2009}, suggesting a mathematical framework could be well received. It may be that the HMM is a good setting to formulate a well principled model of disorder, then test its validity by comparing theoretical and experimental cross sections.

This framework is open to much theoretical development. For example, for a crystal with uncountably many hidden states, the operator $\mathcal{T}$ is a Fredholm integral operator connecting the mathematical theory to the well developed mathematical field of Fredholm theory. Exploring this connection may answer some practical questions, like how the eigenvalues of $\mathcal{T}$ are related to the convergence of the state distribution to steady state, which is discussed for the case of finitely many hidden states by \citeasnoun{riechers2015pairwise}. These authors pose the major problem of chaotic crystallography as reconstructing a crystal's $\varepsilon$-machine from scattering data, where the $\varepsilon$-machine is (roughly speaking) the most information theoretically simple process that could give rise to the observed scattering pattern. Developing a theory of how to construct the $\varepsilon$-machine of a turbostratic material could be of great use to any community studying a turbostratically disordered material.

Moreover, our treatment of countably infinite spaces is brief, and could plausibly be developed into something more readily applicable to crystals with a countably infinite number of layer types like moir\'{e} graphene. Bringing moir\'{e} graphene into the purview of chaotic crystallography could shed some new light on the mysterious material, and be a fruitful area of research.

\begin{appendix}

\section{Powder average}
\label{powder_average_appendix}

To make sense of our expression, consider a Cartesian coordinate system, where the unit length is equal to $2 \pi |\vec{c}^*|$. A point $\vec{Q} = (h,k,l)$ in the reciprocal lattice coordinates will be denoted $\vec{Q}' = (h',k',l')$ in the Cartesian coordinates. Suppose a layer of type $x$ is composed of $n_x$ atoms and the $i$th atom is found at position $\big( a_i(x) , b_i(x), c_i(x) \big)$ in these Cartesian coordinates. Then the square distance between the $i$th atom in a layer of type $x$ and the $j$th atom in a layer of type $y$ for layers $x,y$ separated by $m_3$ is 
\begin{align}
    \big( L^{m_3}_{ij}(x,y) \big)^2 = ( a_i(x) - a_j(y) )^2 + ( b_i(x) - b_j(y) )^2 \\ + ( c_i(x) - c_j(y) + m_3)^2. \nonumber
\end{align}
In the Cartesian system the structure factor of a layer $x$ is
\begin{align}
    F(x) = \sum^{n_x}_{i = 1} f_i e^{i ( a_i(x) h' + b_i(x) k' + c_i(x) l'}
\end{align}
and the structure factor product
\begin{align}
    F(x) F^*(y) = \sum^{n_x}_{i = 1}\sum^{n_y}_{j = 1} f_i f_j e^{i \big( (a_i(x) - a_j(y)) h' + ( b_i(x) - b_j(y) ) k' + ( c_i(x) - c_j(y) ) l' \big)}
\end{align}

Now, to derive our expression for the cross section, we will use the Harmonic expansion presented by \citeasnoun{doi:10.1002/zamm.19860660108} 
\begin{align}
    \frac{1}{4 \pi Q^2}\int_{\partial B_Q} \frac{d \sigma}{d \Omega} dS =
    \sum^{\infty}_{n=0} \frac{Q^{2n}}{(2n+1)!} \Delta^n \bigg( \frac{d \sigma}{d \Omega} \bigg) \bigg\rvert_{\vec{Q}' = 0}
\end{align}
where
\begin{align}
    \Delta^n = \bigg( \frac{\partial^2}{\partial h'^2} + \frac{\partial^2}{\partial k'^2} + \frac{\partial^2}{\partial l'^2} \bigg)^n.
\end{align}
We will begin by writing down the cross section
\begin{align}
    \frac{d \sigma}{d \Omega} = \sum_{m_3 = -N_c}^{N_c} (N_c - m_3) \int_{\mathcal{A}} \int_{\mathcal{A}} G_{m_3}(x,y) F(x)F^*(y) e^{2 \pi i m_3 l} dx dy
\end{align}
and then seek a nice expression for $\Delta^n \big( \frac{d \sigma}{d \Omega} \big) \big\rvert_{\vec{Q}' = 0}$. So we consider
\begin{align}
    \Delta^n \bigg( \frac{d \sigma}{d \Omega} \bigg) \bigg \rvert_{\vec{Q}' = 0} = \sum_{m_3 = -N_c}^{N_c} (N_c - m_3) \int_{\mathcal{A}} \int_{\mathcal{A}} G_{m_3}(x,y) \\
    \times \Delta^n \bigg( F(x)F^*(y) e^{2 \pi i m_3 l} \bigg) \bigg\lvert_{\vec{Q}' = 0} dx dy \nonumber
\end{align}
and bring our attention to
\begin{align}
    \Delta^n \bigg( F(x)F^*(y) e^{2 \pi i m_3 l} \bigg) \bigg\lvert_{\vec{Q}' = 0} \\
    = \sum^{n_x}_{i = 1} \sum^{n_y}_{j = 1} 
    \Delta^n \bigg( f_i f_j \exp\big( i( (a_i(x) &- a_j(y))h' \nonumber \\ + ( b_i(x) &- b_j(y) )k' \nonumber \\ + ( c_i(x) &- c_j(y) + m_3)l' )\big) \bigg) \bigg\rvert_{\vec{Q}'} \nonumber
    \\
    = \sum^{n_x}_{i = 1} \sum^{n_y}_{j = 1} f_i f_j (-1)^n \big( (a_i(x) &- a_j(y) )^2 \\ + ( b_i(x) &- b_j(y) )^2 \nonumber \\ + ( c_i(x) &- c_j(y) + m_3 )^2 \big)^n \nonumber
    \\ =
    \sum^{n_x}_{i = 1} \sum^{n_y}_{j = 1} f_i f_j (-1)^n \big( L^{m_3}_{ij}(x,y) \big)^{2n}. \nonumber
\end{align}
Next, we notice
\begin{align}
    \sum^{\infty}_{n=0}\frac{Q^{2n}}{(2n+1)!} \Delta^n \bigg( F(x)F^*(y) e^{2 \pi i m_3 l} \bigg) \bigg\lvert_{\vec{Q}' = 0} \\ =
    \sum^{\infty}_{n=0}\frac{Q^{2n}}{(2n+1)!}  \sum^{n_x}_{i = 1} \sum^{n_y}_{j = 1} f_i f_j (-1)^n \big( L^{m_3}_{ij}(x,y) \big)^{2n} \nonumber
    \\ =
    \sum^{n_x}_{i = 1} \sum^{n_y}_{j = 1} f_i f_j \sum^{\infty}_{n=0}\frac{( Q L^{m_3}_{ij}(x,y) )^{2n}}{(2n+1)!} (-1)^n \nonumber \\ =
    \sum^{n_x}_{i = 1} \sum^{n_y}_{j = 1} f_i f_j \text{sinc}(Q L^{m_3}_{ij}(x,y))
\end{align}
where we have used the series expansion of $\text{sinc}$
\begin{align}
    \text{sinc}(x) = \sum_{n=0}^{\infty}\frac{x^{2n}}{(2n+1)!}(-1)^n.
\end{align}
Putting all this together
\begin{align}
    \frac{1}{4 \pi Q^2}\int_{\partial B_{Q}} \frac{d \sigma}{d \Omega} dS &= \sum^{\infty}_{n=0} \frac{Q^{2n}}{(2n+1)!} \Delta^n \bigg( \frac{d \sigma}{d \Omega} \bigg) \bigg \rvert_{\vec{Q}' = 0} \\ &=
    \sum_{m_3 = -N_c}^{N_c} (N - m_3) \int_{\mathcal{A}} \int_{\mathcal{A}} G_{m_3}(x,y) \\
    \times \sum^{\infty}_{n=0}\frac{Q^{2n}}{(2n+1)!}& \Delta^n \bigg( F(x)F^*(y) e^{2 \pi i m_3 l} \bigg) \bigg\lvert_{\vec{Q}' = 0} dx dy \nonumber
    \\ &=
    \sum_{m_3 = -N_c}^{N_c} (N_c - m_3) \int_{\mathcal{A}} \int_{\mathcal{A}} G_{m_3}(x,y) \\
    &\times \sum^{n_x}_{i = 1} \sum^{n_y}_{j = 1} f_i f_j \text{sinc}(Q L^{m_3}_{ij}(x,y)) dx dy \nonumber.
\end{align}

\section{Finite state space, uncountable alphabet}
\label{FSSUA}

Suppose we have a finite state space $\mathbb{S}$ and alphabet $\mathcal{A} \subset \mathbb{R}^n$ an open connected set. Then by combining the expression for the pair correlation function \eqref{Pair_Correlation_Finite_S} with the average structure factor product \eqref{Average_structure_factor_product} we arrive at

\begin{align}
    Y_m = \int_{\mathcal{A}}\int_{\mathcal{A}}\sum_{r \in \mathbb{S}}\sum_{s \in \mathbb{S}} \pi_s \mathcal{T}^m_{sr}v_{r}(x)F(x)F^*(y)v_s (y) dx dy \\ =
    \sum_{r \in \mathbb{S}}\sum_{s \in \mathbb{S}} \pi_s \mathcal{T}^m_{sr} \int_{\mathcal{A}} v_r(x)F(x) dx \int_{\mathcal{A}} v_s(y)F^*(y) dy \\ =
    \sum_{r \in \mathbb{S}}\sum_{s \in \mathbb{S}} \pi_s \mathcal{T}^m_{sr} h_{rs} \\ =
    \Tr \big( \Diag(\pi)\mathcal{T}^m H \big)
\end{align}
where $H$ is a Hermitian matrix with dimension equal to the number of states in the space $\mathbb{S}$, and has elements
\begin{align}
    h_{rs} = \int_{\mathcal{A}} v_r(x)F(x) dx \int_{\mathcal{A}} v_s(y)F^*(y) dy.
\end{align}
Next, we note that
$Y_m = Y_{-m}^*$ so 
\begin{align}
    \frac{d \sigma}{d \Omega} = \sum_{m_3 =-N_c}^{N _c}(N_c - \abs{m_3})Y_{m_3} e^{2 \pi i m_3 l} \\ =
    \sum_{m_3 = 1}^{N _c}(N_c - m_3)Y_{m_3} e^{2 \pi i m_3 l} \\ +
    N_c Y_0 \nonumber \\ + 
    \sum_{m_3 = 1}^{N _c}(N_c - m_3)Y_{m_3}^* e^{-2 \pi i m_3 l} \nonumber \\ =
    \sum_{m_3 = 1}^{N _c}(N_c - m_3)\Tr \big( \Diag(\pi)\mathcal{T}^m H \big)e^{2 \pi i m_3 l} \\ +
    N_c \Tr \big( \Diag(\pi) H \big) \nonumber \\ + 
    \sum_{m_3 = 1}^{N _c}(N_c - m_3)\Tr \big( \Diag(\pi)\mathcal{T}^m H^* \big)e^{- 2 \pi i m_3 l} \nonumber \\ =
        \Tr \Bigg( \Diag(\pi)\bigg(\sum_{m_3 = 1}^{N _c}(N_c - m_3)(\mathcal{T}e^{2 \pi i l})^{m_3}\bigg) H \Bigg) \\ +
    N_c \Tr \big( \Diag(\pi) H \big) \nonumber \\ + 
    \Tr \Bigg( \Diag(\pi)\bigg(\sum_{m_3 = 1}^{N _c}(N_c - m_3)(\mathcal{T}e^{-2 \pi i l})^{m_3}\bigg) H^* \Bigg) \nonumber \\ =
            \Tr \big( \Diag(\pi)S H \big) \\ +
    N_c \Tr \big( \Diag(\pi) H \big) \nonumber \\ + 
    \Tr \big( \Diag(\pi)S^* H^* \big) \nonumber \\ =
    2 Re \bigg\{\Tr \big( \Diag(\pi)H(2S + N_c I) \big) \bigg\}
\end{align}
and we arrive at the expression for cross section.

\section{Infinite state space}
\label{UncountableAppendix}

This section describes a crystal comprising infinitely many layer types, where the probability distribution over the set of layers follows a Markov chain. In some cases, the arguments for an uncountable space represented by $\mathbb{R}^n$ (open and connected) and a countable space represented by $\mathbb{N}$ are essentially the same, and in these cases arguments may be made over a general Hilbert space $\mathcal{H}$ that apply to both the square summable sequences $\ell^2$ and the square integrable functions $L^2$, representing distributions over the countable state space $\mathbb{N}$ or uncountable state space $\mathbb{R}^n$ (open and connected) respectively.
For brevity, we allow the symbols of integration
\begin{align}
    \int \cdot \ dx
\end{align}
to represent either integration over the open connected set $\mathcal{A} \subset \mathbb{R}^n$ or summation over $\mathbb{N}$. 

\subsection{The most general kernel}
\label{general_kernel}

The cross section of a crystal with a countable infinity or uncountable infinity of hidden layers is
\begin{align}
\sum_{m_{3} = -N_{c}}^{N_{c}}(N_c - \abs{m_3}) \int \int F(x)F^*(y)G_{m_3}(x,y)e^{2 \pi i m_3 l} dx dy, \nonumber
\end{align}
which we can split into three terms
\begin{align}
&\sum_{m_{3} = -N_{c}}^{N_{c}}(N_c - \abs{m_3}) \int \int F(x)F^*(y)G_{m_3}(x,y)e^{2 \pi i m_3 l} dx dy \label{simpUncCS} \\
=&\sum_{m_{3} = 1}^{N_{c}}(N_c - \abs{m_3}) \int \int F(x)F^*(y)G_{m_3}(x,y)e^{2 \pi i m_3 l} dx dy \nonumber \\
+& N_c \int \int F(x)F^*(y)G_{0}(x,y)dx dy \nonumber \\
+&\sum_{m_{3} = 1}^{N_c}(N_c - \abs{m_3}) \int \int F(x)F^*(y)G_{-m_3}(x,y)e^{-2 \pi i m_3 l} dx dy. \nonumber
\end{align}
Now the second term of the RHS of equation \eqref{simpUncCS} requires an expression for $G_{0}(x,y)$, which represents the probability (density) of sampling a layer that is type $x$, and given it is type $x$ that it is itself type $y$. In an uncountable space we let $\delta_a(y) \equiv \delta(a-y)$ for $a \in \mathcal{A}$ represent the shifted Dirac delta function evaluated at $y \in \mathcal{A}$ and notice $G_{0}(x,y) = \pi(x)\delta_x(y)$. In a countable space, $G_{0}(i,j) = \pi_i\delta_{ij}$ for $ij \in \mathbb{N}$ by the same arguments. Again, to avoid writing essentially the same thing twice, we continue by allowing $\delta_x(y)$ to represent $\delta_{xy}$ for countable spaces. With this, we have that the second term satisfies
\begin{align}
& N_c \int \int F(x)F^*(y)G_{0}(x,y)dx dy \\
=& N_c \int \pi(x) F^*(x) F(x) dx. \nonumber
\end{align} 
Next, we focus on the third term, noting the probability (density) of sampling a state $x$ then finding a state $y$ after moving forward $m_3$ blocks is the same as sampling a state $y$ then finding a state $x$ after moving backward $m_3$ blocks. Thus
\begin{equation}
F(x)F^*(y) G_{m_3}(x,y) = F^*(x) F(y) G_{-m_3}(x,y)
\end{equation}
so we have that the LHS of \eqref{simpUncCS} equals
\begin{align}
&\sum_{m_{3} = 1}^{N_{c}}(N_c - \abs{m_3}) \int \int F(x)F^*(y)G_{m_3}(x,y)e^{2 \pi i m_3 l} dx dy \\
+& N_c \int \pi(x) F^*(x) F(x) dx \nonumber \\
+&\sum_{m_{3} = 1}^{N_c}(N_c - \abs{m_3}) \int \int F^*(x) F(y)G_{m_3}(x,y)e^{-2 \pi i m_3 l} \nonumber dx dy
\end{align}
hence we can see that the third term is just the complex conjugate of the first. With this information, note that the sum of the first and third term is just twice the real part of the first, so we proceed by only considering the first term and noting
\begin{align}
    G_{m_3}(x,y) \equiv \pi(x)\mathcal{T}^{m_3}\delta_x(y) \text{ for $m_3 > 0$} 
\end{align}
so
\begin{align}
&\sum_{m_{3} = 1}^{N_{c}}(N_c - \abs{m_3}) \int \int F(x)F^*(y)G_{m_3}(x,y)e^{2 \pi i m_3 l} dx dy \label{uncountableCS} \\
=&\int\int F(x)F^*(y) \pi(x) \sum_{m_{3} = 1}^{N_c}(N_c - \abs{m_3})(e^{2 \pi i l}\mathcal{T})^{m_3} \delta_x(y) dx dy \nonumber \\
=& \int\int F(x)F^*(y)\pi(x)Z\delta_x(y) dx dy
\end{align}
where $z \in \mathbb{C}$ and $Z$ is an operator with expression 
\begin{align}
Z &\equiv \sum_{m_{3} = 1}^{N_c}(N_c - \abs{m_3})(e^{2 \pi i l}\mathcal{T})^{m_3}.
\end{align}
Putting all this together, we have
\begin{align}
\frac{d \sigma}{d \Omega} =
2 Re \Bigg\{ \int\int F(x)F^*(y)\pi(x)Z\delta_x(y) dx dy \Bigg\}  \\
+ N_c \int \abs{F(x)}^2 \pi(x) dx \nonumber
\end{align}
where $Re\{z\}$ represents the real part of the complex number $z \in \mathbb{C}$. This completes the derivation of the general cross section for both countable and uncountable state spaces.

\subsection{The case of a symmetric kernel}
\label{self_adj_app}

With the general expression established for both countable and uncountable spaces, we consider the special case where $k(x,y) = k(y,x)$ so $\mathcal{T}$ is self-adjoint. Since $\mathcal{T}$ is compact, we observe by the Spectral Theorem for compact self-adjoint operators that
\begin{align}
(e^{2 \pi i l}\mathcal{T})^{m_3} \delta_x(y) &= \sum_{n \in \Lambda} \langle \delta_x , u_n \rangle (\lambda_n e^{2 \pi i l})^{m_3} u_{n}(y) \\
&= \sum_{n \in \Lambda} u_n(x) u_{n}(y) (\lambda_ne^{2 \pi i l})^{m_3} 
\end{align}
where $u_n$ and $\lambda_n e^{2 \pi i l}$ are the eigenvectors and eigenvalues of $e^{2 \pi i l}\mathcal{T}$ indexed by the set $\Lambda$ which repeats eigenvalues according to their algebraic multiplicity. Here we have also used $\langle \phi , \psi \rangle$ to denote inner product on the Hilbert space $\mathcal{H}$ of $\phi, \psi \in \mathcal{H}$. With this, we can proceed from equation \eqref{uncountableCS} and deduce
\begin{align}
&\int\int F(x)F^*(y) \pi(x) \times \\ 
&\sum_{m_{3} = 1}^{N_c}(N_c - \abs{m_3})(e^{2 \pi i l}\mathcal{T})^{m_3} \delta_x(y) dx dy \nonumber \\
= &\int \int F(x)F^*(y) \pi(x) \times \nonumber \\
 &\sum_{m_{3} = 1}^{N_c}(N_c - \abs{m_3})\sum_{n \in \Lambda} u_n(x) u_{n}(y) (\lambda_n e^{2 \pi i l})^{m_3} dx dy \nonumber \\
= &\sum_{n \in \Lambda}\int F(x) \pi(x) u_n(x) \int  F^*(y) u_n(y) \times  \\ 
&\sum_{m_{3} = 1}^{N_c}(N_c - \abs{m_3}) (\lambda_n e^{2 \pi i l})^{m_3} dx dy \nonumber \\
= &\sum_{n \in \Lambda} s_n\int F(x) \pi(x) u_n(x) dx \int  F^*(y) u_n(y)  dy \label{uncountableSA}
\end{align}
where
\begin{align}
s_{n} = 
\begin{cases}
\frac{N_c}{2}(N_c - 1) \text{ if $l \in \mathbb{Z}$ and $\lambda_n = 1$ }  \\
\frac{\lambda_n e^{2 \pi i l} ( \lambda_n^{N_c} e^{2 \pi i l N_c} + N_c(1 - \lambda_n e^{2 \pi i l} ) - 1)}{(1 - \lambda_n e^{2 \pi i l})^2} \text{ otherwise,}
\end{cases}
\end{align}
and the expression for the cross section follows immediately.
We can evaluate \eqref{uncountableSA} approximately by summing over only the first few values of $n$, requiring only a few eigenvalues and eigenvectors. Unfortunately, there is no method of deriving closed form solutions for $u_n$ for a general kernel $k$; but analytic solutions do exist in some special cases. In the case of an uncountable space represented by $\mathcal{A}$ it may be fruitful to note that solutions to the eigenvalue equation
\begin{align}
\lambda_n u_n(x) = \int_{\mathcal{A}}k(x,y)u_n(y)dy
\end{align}
also satisfy
\begin{align}
Lu_n = \lambda_n^{-1} u_n \label{DE}
\end{align}
for $L$ a differential operator with kernel $k$. Finding the eigenvalues and eigenvectors of $L$ is then a question of solving the differential equation \eqref{DE}.

\subsection{The case of a convolution kernel}
\label{conv_ker_appendix}

Let $\mathcal{A} = (0,1)^n$, $\phi \in L^2$ and
\begin{align}
    k(x,y) = \sum_{m \in \mathbb{Z}^n} P(m + y-x)
\end{align}
where $P$ is a square integrable probability distribution over $\mathbb{R}^n$. Then
\begin{align}
\mathcal{T} \phi =
\int_{\mathcal{A}}k(x,y)\phi(x) dx \\ = 
\sum_{m \in \mathbb{Z}^n}\int_{\mathcal{A}}P(m+y-x)\phi(x) dx \\ =
\int_{\mathbb{R}^n}P(y-x)\phi(x) dx \equiv P \circledast \phi
\end{align}
where $\circledast$ denotes the convolution. Now define the sequence of functions
\begin{align}
\mathcal{T}^{n+1}\phi = P \circledast \mathcal{T}^{n} \phi
\end{align}
for which we denote the $n$th term by $P\circledast^{n} \phi$, and identify this sequence with $\mathcal{T}^n \phi$.
Next, we denote the Fourier transform by $\mathcal{F}: L^2 \to L^2$ and use the convolution theorem to deduce
\begin{align}
\mathcal{F} \big[ P\circledast^{n} \phi \big] = (\mathcal{F} [P])^n\mathcal{F}[\phi].
\end{align}
We now observe that
\begin{align}
Z \delta_x &= \sum_{m_{3} = 1}^{N_c}(N_c - \abs{m_3})(e^{2 \pi i l}\mathcal{T})^{m_3}\delta_x\\
&= \sum_{m_{3} = 1}^{N_c}(N_c - \abs{m_3}) e^{2 \pi i m_3 l}P \circledast^{m_3}\delta_x\\
&= \mathcal{F}^{-1}\bigg[\sum_{m_{3} = 1}^{N_c}(N_c - \abs{m_3}) e^{2 \pi i m_3 l}\mathcal{F}\big[P \circledast^{m_3}\delta_x\big] \bigg]\\
&= \mathcal{F}^{-1}\bigg[\sum_{m_{3} = 1}^{N_c}(N_c - \abs{m_3}) e^{2 \pi i m_3 l}\mathcal{F}[P]^{m_3}\mathcal{F}[\delta_x] \bigg] \\
&= \mathcal{F}^{-1}\bigg[\mathcal{F}[s]\mathcal{F}[\delta_x] \bigg] \\
&= s \circledast \delta_x \\
&= s_x
\end{align}
where $s \in L^2$ is defined
\begin{align}
s \equiv \mathcal{F}^{-1}\bigg[ \sum_{m_{3} = 1}^{N_c}(N_c - \abs{m_3}) e^{2 \pi i m_3 l}\mathcal{F}[P]^{m_3}\bigg]
\end{align}
so satisfies
\begin{align}
\mathcal{F}[s] = 
\begin{cases}
\frac{N_c}{2}(N_c - 1) \text{ if $\mathcal{F}[P]e^{2 \pi i l} = 1$}  \\
\frac{\mathcal{F}[P] e^{2 \pi i l} ( \mathcal{F}[P]^{N_c} e^{2 \pi i l N_c} + N_c(1 - \mathcal{F}[P] e^{2 \pi i l} ) - 1)}{(1 - \mathcal{F}[P] e^{2 \pi i l})^2} \text{ otherwise,}
\end{cases}
\end{align}
and $s_x(y) \equiv s(y-x)$
allowing us to arrive at an expression for the cross section
\begin{align}
\frac{d \sigma}{d \Omega} =
2 Re \Bigg\{ \int_{\mathcal{A}}\int_{\mathbb{R}^n}F(x)F^*(y)s(y-x) dx dy \Bigg\} \label{conv_ker_cs} \\
+ N_c \int_{\mathcal{A}} \abs{F(x)}^2 dx \nonumber \\
= Re \Bigg\{ \int_{\mathcal{A}}2F^*(x)(F \circledast s)(x) dy 
+ N_c \abs{F(x)}^2 dx \Bigg\}.
\end{align}
If we make the further assumption that layers are identical up to translation, then starting from equation \eqref{conv_ker_cs}
\begin{align}
    \frac{d \sigma}{d \Omega} = 2 Re \bigg\{ \int_{\mathcal{A}}\int_{\mathbb{R}^n} F(x)F^*(y)s(y-x) dx dy \bigg\} \\ + N_c \int_\mathcal{A}\abs{F(x)}^2 dx \nonumber \\ = 2 Re \bigg\{ \abs{F(0)}^2\int_{\mathcal{A}}\int_{\mathbb{R}^n} e^{2 \pi i \vec{x}\cdot\vec{Q}}e^{-2 \pi i \vec{y}\cdot\vec{Q}} s(y-x) dx dy \bigg\} \\ + N_c \abs{F(0)}^2\int_\mathcal{A} dx \nonumber \\
    = 2 Re \bigg\{ \abs{F(0)}^2\int_{\mathcal{A}} e^{2 \pi i \vec{x}\cdot\vec{Q}}\int_{\mathbb{R}^n}e^{-2 \pi i \vec{y}\cdot\vec{Q}} s(y-x) dy dx \bigg\} \\ + N_c \abs{F(0)}^2 \nonumber \\
    = 2 Re \bigg\{ \abs{F(0)}^2\int_{\mathcal{A}} e^{2 \pi i \vec{x}\cdot\vec{Q}} \mathcal{F}[s](\vec{Q})e^{- 2 \pi i \vec{x}\cdot\vec{Q}} dx \bigg\} \\ + N_c \abs{F(0)}^2 \nonumber \\
    = 2 Re \bigg\{ \abs{F(0)}^2 \mathcal{F}[s](\vec{Q}) \int_{\mathcal{A}}dx \bigg\} \\ + N_c \abs{F(0)}^2 \nonumber \\
    = \abs{F(0)}^2 \bigg( 2 Re\big\{ \mathcal{F}[s](\vec{Q}) \big\} + N_c\bigg)
\end{align}
and we arrive at the expression for the cross section of a crystal with convolution kernel composed of layers that are identical up to translation.
\end{appendix}

\bibliographystyle{iucr}
\bibliography{References}

@article{PhysRevB.34.3586,
  Title = {Effect of stacking faults on diffraction: the structure of lithium metal},
  Author = {Berliner, R. and Werner, S. A.},
  Journal = {Phys. Rev. B},
  Volume = {34},
  Issue = {6},
  Pages = {3586--3603},
  Numpages = {0},
  Year = {1986},
  Month = {Sep},
  Publisher = {American Physical Society},
  doi = {10.1103/PhysRevB.34.3586},

}

@article{0953-8984-20-28-285105,
  Author={T C Hansen and M M Koza and P Lindner and W F Kuhs},
  Title={Formation and annealing of cubic ice: 2. Kinetic study},
  Journal={Journal of Physics: Condensed Matter},
  Volume={20},
  Number={28},
  Pages={285105},
  Year={2008},
  }

@book{Disorder_first_second,
	Author = {Victor A Dritz and Cyril Tchoubar},
	Publisher = {Springer},
	Title = {X-ray Diffraction by Disordered Lamellar Structures},
	Year = {1990},
	Pages = {19}}

@book{Guinier1964,
	Author = {A Guinier},
	Publisher = {Springer},
	Title = {Th{\'e}orie et technique de la radiocristallographie},
	Year = {1964}}

@article{Hansen2008,
	Author = {T Hansen and M Koza and W F Kuhs},
	Journal = {J. Phys.: Condens. Matter },
	Month = {June},
	Title = {Formation and annealing of cubic ice: 1. Modelling and stacking faults},
	Volume = {20},
	Pages={285104},
	Year = {2008}}

@book{Warren_text,
	Author = {B E Warren},
	Publisher = {Addison-Wesley},
	Title = {X-ray Diffraction},
	Year = {1969}}

@article{Varn201547,
title = "Chaotic crystallography: how the physics of information reveals structural order in materials ",
author={D P Varn and J P Crutchfield},
journal = "Current Opinion in Chemical Engineering ",
volume = "7",
number = "",
pages = "47 - 56",
year = "2015",
note = "",
issn = "2211-3398",
doi = "http://dx.doi.org/10.1016/j.coche.2014.11.002",
}

@inproceedings{wilson1942imperfections,
  title={Imperfections in the structure of cobalt. II. Mathematical treatment of proposed structure},
  author={Wilson, AJ Cl},
  booktitle={Proceedings of the Royal Society of London A: Mathematical, Physical and Engineering Sciences},
  volume={180},
  pages={277--285},
  year={1942},
  organization={The Royal Society}
}

@article{varn2013machine,
  title={$\varepsilon$-Machine spectral reconstruction theory: a direct method for inferring planar disorder and structure from X-ray diffraction studies},
  author={Varn, DP and Canright, GS and Crutchfield, JP},
  journal={Acta Crystallographica Section A: Foundations of Crystallography},
  volume={69},
  number={2},
  pages={197--206},
  year={2013},
  publisher={International Union of Crystallography}
}

@article{warren1941x,
  title={X-ray diffraction in random layer lattices},
  author={Warren, BE},
  journal={Physical Review},
  volume={59},
  number={9},
  pages={693},
  year={1941},
  publisher={APS}
}

@article{riechers2015pairwise,
  title={Pairwise correlations in layered close-packed structures},
  author={Riechers, Paul M and Varn, Dowman P and Crutchfield, James P},
  journal={Acta Crystallographica Section A: Foundations and Advances},
  volume={71},
  number={4},
  pages={423--443},
  year={2015},
  publisher={International Union of Crystallography}
}

@article{varn2016did,
  title={What did Erwin mean? The physics of information from the materials genomics of aperiodic crystals and water to molecular information catalysts and life},
  author={Varn, Dowman P and Crutchfield, James P},
  journal={Phil. Trans. R. Soc. A},
  volume={374},
  number={2063},
  pages={20150067},
  year={2016},
  publisher={The Royal Society}
}

@article{Bistritzer26072011,
author = {Bistritzer, Rafi and MacDonald, Allan H.}, 
title = {Moire bands in twisted double-layer graphene},
volume = {108}, 
number = {30}, 
pages = {12233-12237}, 
year = {2011}, 
doi = {10.1073/pnas.1108174108}, 
eprint = {http://www.pnas.org/content/108/30/12233.full.pdf}, 
journal = {Proceedings of the National Academy of Sciences} 
}

@article{PhysRevLett.99.256802,
  title = {Graphene Bilayer with a Twist: Electronic Structure},
  author = {Lopes dos Santos, J. M. B. and Peres, N. M. R. and Castro Neto, A. H.},
  journal = {Phys. Rev. Lett.},
  volume = {99},
  issue = {25},
  pages = {256802},
  numpages = {4},
  year = {2007},
  month = {Dec},
  publisher = {American Physical Society},
  doi = {10.1103/PhysRevLett.99.256802}
}

@Article{Huang2017,
author={Huang, Shengqiang
and Yankowitz, Matthew
and Chattrakun, Kanokporn
and Sandhu, Arvinder
and LeRoy, Brian J.},
title={Evolution of the electronic band structure of twisted bilayer graphene upon doping},
journal={Scientific Reports},
year={2017},
volume={7},
number={1},
pages={7611},
issn={2045-2322},
doi={10.1038/s41598-017-07580-3}
}

@Article{Razado-Colambo2016,
author={Razado-Colambo, I.
and Avila, J.
and Nys, J.-P.
and Chen, C.
and Wallart, X.
and Asensio, M.-C.
and Vignaud, D.},
title={NanoARPES of twisted bilayer graphene on SiC: absence of velocity renormalization for small angles},
journal={Scientific Reports},
year={2016},
month={Jun},
day={06},
volume={6},
pages={27261}
}

@Article{Ellison2009,
author="Ellison, Christopher J.
and Mahoney, John R.
and Crutchfield, James P.",
title="Prediction, Retrodiction, and the Amount of Information Stored in the Present",
journal="Journal of Statistical Physics",
year="2009",
month="Sep",
day="22",
volume="136",
number="6",
pages="1005",
issn="1572-9613",
doi="10.1007/s10955-009-9808-z"
}

@article{doi:10.1021/nl301137k,
author = {Havener, Robin W. and Zhuang, Houlong and Brown, Lola and Hennig, Richard G. and Park, Jiwoong},
title = {Angle-Resolved Raman Imaging of Interlayer Rotations and Interactions in Twisted Bilayer Graphene},
journal = {Nano Letters},
volume = {12},
number = {6},
pages = {3162-3167},
year = {2012},
doi = {10.1021/nl301137k},
    note ={PMID: 22612855},
    
eprint = { 
        http://dx.doi.org/10.1021/nl301137k
    
}

}

@article{doi:10.1021/nl204547v,
author = {Brown, Lola and Hovden, Robert and Huang, Pinshane and Wojcik, Michal and Muller, David A. and Park, Jiwoong},
title = {Twinning and Twisting of Tri- and Bilayer Graphene},
journal = {Nano Letters},
volume = {12},
number = {3},
pages = {1609-1615},
year = {2012},
doi = {10.1021/nl204547v},
    note ={PMID: 22329410},

eprint = { 
        http://dx.doi.org/10.1021/nl204547v
    
}

}

@article{LI20071686,
title = "X-ray diffraction patterns of graphite and turbostratic carbon",
journal = "Carbon",
volume = "45",
number = "8",
pages = "1686 - 1695",
year = "2007",
issn = "0008-6223",
doi = "https://doi.org/10.1016/j.carbon.2007.03.038",
author = "Z.Q. Li and C.J. Lu and Z.P. Xia and Y. Zhou and Z. Luo"
}

@article {Ufer:2008:0009-8604:272,
title = "Quantitative phase analysis of bentonites by the Rietveld method",
journal = "Clays and Clay Minerals",
parent_itemid = "infobike://cms/ccm",
publishercode ="cms",
year = "2008",
volume = "56",
number = "2",
pages = "272-282",
itemtype = "ARTICLE",
issn = "0009-8604",
doi = "doi:10.1346/CCMN.2008.0560210",
author = "Ufer, K. and Stanjek, H. and Roth, G. and Dohrmann, R. and Kleeberg, R. and Kaufhold, S.",
}

@article{Turbostratic2009,
title = "Description of X-ray powder pattern of turbostratically disordered layer structures with a Rietveld compatible approach",
journal = "Zeitschrift f{\"u}r Kristallographie - Crystalline Materials",
volume = "219",
number = "9",
pages = "519-527",
year = "2009",
author = "Kristian Ufer and Georg Roth and Reinhard Kleeberg and Helge Stanjek and Reiner Dohrmann and Jörg Bergmann"
}

@article{UNGAR2002929,
title = "Microstructure of carbon blacks determined by X-ray diffraction profile analysis",
journal = "Carbon",
volume = "40",
number = "6",
pages = "929 - 937",
year = "2002",
author = "Tam\'{a}s Ung\'{a}r and Jen{\"o} Gubicza and G\'{a}bor Rib\'{a}rik and Cristian Pantea and T Waldek Zerda"
}

@article{ZHOU201417,
title = "Interpretation of X-ray diffraction patterns of (nuclear) graphite",
journal = "Carbon",
volume = "69",
pages = "17 - 24",
year = "2014",
issn = "0008-6223",
doi = "https://doi.org/10.1016/j.carbon.2013.11.032",
author = "Z. Zhou and W.G. Bouwman and H. Schut and C. Pappas"
}

@article{Shihw0018,
title = "Structure-refinement program for disordered carbons",
journal = "Journal of Applied Crystallography",
volume = "26",
number = "6",
pages = "827 - 836",
year = "1993",
author = "Shi, H. and Reimers, J. N. and Dahn, J. R."
}

@phdthesis{ShiThesis,
    title    = "Disordered Carbons and Battery Applications",
    school   = "Simon Fraser University",
    author   = "Shi, Hang",
    year     = "1993",
}

@Article{superconducting_angle,
author={Yuan Cao and Valla Fatemi and Shiang Fang and Kenji Watanabe and Takashi Taniguchi and Efthimios Kaxiras and Pablo Jarillo-Herrero},
title={Unconventional superconductivity in magic-angle graphene superlattices},
journal={Nature},
year={2018},
month={April},
day={05},
pages={43-50},
volume={556},
publisher={Macmillan Publishers Limited, part of Springer Nature. All rights reserved. SN  -}
}

@article{ERGUN1976139,
title = "Analysis of coherence, strain, thermal vibration and preferred orientation in carbons by X-Ray diffraction",
journal = "Carbon",
volume = "14",
number = "3",
pages = "139 - 150",
year = "1976",
issn = "0008-6223",
doi = "https://doi.org/10.1016/0008-6223(76)90094-4",
author = "Sabri Ergun"
}

@article{MarkovPaper1,
  title={A Markov theoretic description of stacking disorder and aperiodic crystals},
  author={A G Hart and T C Hansen and W F Kuhs},
  journal={Acta Crystallographica Section A: Foundations of Crystallography},
  volume={74},
  pages={357--372},
  year={2018},
  publisher={International Union of Crystallography}
}

@article{Warren1942,
author = {Biscoe,J.  and Warren,B. E. },
title = {An X‐Ray Study of Carbon Black},
journal = {Journal of Applied Physics},
volume = {13},
number = {6},
pages = {364-371},
year = {1942},
doi = {10.1063/1.1714879},
eprint = { 
        https://doi.org/10.1063/1.1714879
    
}

}

@article{doi:10.1002/zamm.19860660108,
author = {Ba{\u{z}}ant, P. and Oh, B. H.},
title = {Efficient Numerical Integration on the Surface of a Sphere},
journal = {ZAMM - Journal of Applied Mathematics and Mechanics / Zeitschrift f{\"{u}}r Angewandte Mathematik und Mechanik},
volume = {66},
number = {1},
pages = {37-49},
year = {1986}
}

@article{Scardi:me0628,
author = "Scardi, Paolo and Billinge, Simon J. L. and Neder, Reinhard and Cervellino, Antonio",
title = "{Celebrating 100 years of the Debye scattering equation}",
journal = "Acta Crystallographica Section A",
year = "2016",
volume = "72",
number = "6",
pages = "589--590",
month = "Nov",
doi = {10.1107/S2053273316015680},
}

@article{doi:10.1107/S0365110X51001409,
author = {Brindley, G. W. and Méring, J.},
title = {Diffractions des rayons X par les structures en couches désordonnées. I},
journal = {Acta Crystallographica},
volume = {4},
number = {5},
pages = {441-447},
year = {1951}
}

\end{document}